\documentclass{article}

\usepackage{arxiv}

\usepackage[T1]{fontenc}    % use 8-bit T1 fonts

\usepackage{nicefrac}       
\usepackage{microtype}      
\usepackage{lipsum}

\usepackage{natbib}
\usepackage{doi}
\usepackage{authblk}
\usepackage{silence}
\usepackage{lineno,hyperref}

\usepackage[utf8]{inputenc}
\usepackage{graphicx}
\usepackage{amsmath}
\usepackage{scrextend}
\usepackage{caption}
\usepackage{subcaption}
\usepackage[table]{xcolor}
\usepackage{float}
\usepackage{paralist}
\usepackage{url}
\usepackage{tabto}
\usepackage{latexsym}
\usepackage{textcomp}
\usepackage{amsfonts}
\usepackage{hyphenat}
\usepackage{amssymb}
\usepackage{multirow}
\usepackage{algorithm}
\usepackage[noend]{algpseudocode}
\usepackage{makecell}
\usepackage{verbatim}
\usepackage{listings}
\lstdefinestyle{sparql}{
    basicstyle=\scriptsize,
    keywordstyle=\bfseries,
    morekeywords={PREFIX, SELECT, DISTINCT, UNION, OPTIONAL, FILTER, WHERE, SERVICE}
}
\hypersetup{
    colorlinks=true,
    linkcolor=blue,
    citecolor=blue,      
    urlcolor=blue
}
\usepackage{xcolor}
\usepackage{adjustbox}
\usepackage[utf8]{inputenc}
\usepackage{booktabs}
\newcommand{\q}[1]{``#1''}

\modulolinenumbers[5]
\begin{document}

%\begin{frontmatter}

\title{Knowledge4COVID-19: A Semantic-based Approach for Constructing a COVID-19 related Knowledge Graph from Various Sources and Analysing Treatments' Toxicities}
%\tnotetext[mytitlenote]{Knowledge4COVID-19}

\author[1,2]{Ahmad Sakor}
\author[1,2]{Samaneh Jozashoori}
\author[1,2]{Emetis Niazmand}
\author[1,2]{Ariam Rivas}
\author[3,4]{Kostantinos Bougiatiotis}
\author[3,*]{Fotis Aisopos\thanks{fotis.aisopos@iit.demokritos.gr}}
\author[1,2]{Enrique Iglesias}
\author[1,2]{Philipp D. Rohde}
\author[1,2]{Trupti Padiya}
\author[3]{Anastasia Krithara}
\author[3]{Georgios Paliouras}
\author[1,2,*]{Maria-Esther Vidal\thanks{Maria.Vidal@tib.eu}}

\affil[1]{TIB Leibniz Information Centre for Science and Technology , Welfengarten 1 B, Hannover, Germany}
\affil[2]{L3S Research Center, University of Hannover, Appelstraße 9a, Hannover, Germany}
\affil[3]{Institute of Informatics \& Telecommunications, NCSR Demokritos. Patr. Grigoriou \& Neapoleos Str, Ag. Paraskevi, Athens, Greece}
\affil[4]{Department of Informatics and Telecommunications, National and Kapodistrian University of Athens, Panepistimiou 30, Athens, Greece}
\date{}                     %% if you don't need date to appear
\setcounter{Maxaffil}{0}
\renewcommand\Affilfont{\itshape\small}

\maketitle  

\begin{abstract}
In this paper, we present Knowledge4COVID-19, a framework that aims to showcase the power of integrating disparate sources of knowledge to discover adverse drug effects caused by drug-drug interactions among COVID-19 treatments and pre-existing condition drugs. 
Initially, we focus on constructing the Knowledge4COVID-19 knowledge graph (KG) from the declarative definition of mapping rules using the RDF Mapping Language.
Since valuable information about drug treatments, drug-drug interactions, and side effects is present in textual descriptions in scientific databases (e.g., DrugBank) or in scientific literature (e.g., the CORD-19, the Covid-19 Open Research Dataset), the Knowledge4COVID-19 framework implements Natural Language Processing.
The Knowledge4COVID-19 framework extracts relevant entities and predicates that enable the fine-grained description of COVID-19 treatments and the potential adverse events that may occur when these treatments are combined with treatments of common comorbidities, e.g., hypertension, diabetes, or asthma.
Moreover, on top of the KG, several techniques for the discovery and prediction of interactions and potential adverse effects of drugs have been developed with the aim of suggesting more accurate treatments for treating the virus.
We provide services to traverse the KG and visualize the effects that a group of drugs may have on a treatment outcome. 
Knowledge4COVID-19 was part of the Pan-European \emph{hackathon\#EUvsVirus} in April 2020 and is publicly available as a resource through a \href{https://github.com/SDM-TIB/Knowledge4COVID-19}{GitHub repository} and a \href{https://zenodo.org/record/4701817#.YH336-8zbol}{DOI}.

\end{abstract}


%\end{frontmatter}

%\linenumbers

\section{Introduction}
In early December 2019, an outbreak of a novel virus, the severe acute respiratory syndrome coronavirus 2 (SARS-CoV-2) occurred in China, causing a rapid spread of the coronavirus disease 2019 (COVID-19). SARS-CoV-2 can be transmitted during the asymptomatic phase of infection and poses a global health emergency because of the intricacy of tracing mild or presymptomatic phases. The disease spectrum of SARS-CoV-2 infection varies in severity from asymptomatic to mild respiratory tract infection and severe or fatal pneumonia. 
The virus infection landscape poses serious challenges that have to be addressed by the research community to come up with the tools that efficiently combat the pandemic. Specifically, the aggregation of heterogeneous data (e.g., publications and open scientific databases) into a common knowledge base will enable the development of data-driven tools. Moreover, data governance, interoperability and data quality issues, and efficient query processing and data exploration are relevant challenges demanded to be solved efficiently. More importantly, it is crucial to explore adverse effects of the treatments commonly prescribed for pre-existing conditions and the potential treatments for COVID-19. 
\\
\textbf{Our Resource:} We address the problem of data integration and propose a resource named Knowledge4COVID-19, which transforms COVID-19 and SARS-CoV-2 related data into a KG. The Knowledge4COVID-19 resource is composed of a data ecosystem (DE) and the Knowledge4COVID-19 KG, both allow for a unified view of the data sources in terms of the unified schema. The different components of the Knowledge4COVID-19 DE enable entity extraction and linking, data curation, and the resolution of the heterogeneity conflicts across the data sources. Moreover, they facilitate the integration of the heterogeneous data into a uniform view. Mapping rules expressed in the RDF mapping language (RML) describe these correspondences~\cite {DimouSCVMW14}. In addition, knowledge extraction methods make use of knowledge encoded in diverse sources for extracting drug-drug interactions. These data sources include controlled vocabularies (e.g., Unified Medical Language System-UMLS\footnote{\url{https://www.nlm.nih.gov/research/umls/index.html}}), scientific publications (e.g., CORD-19\footnote{\url{https://www.semanticscholar.org/cord19}}) and scientific open databases (e.g., DrugBank\footnote{\label{goDrugBank}\url{https://go.DrugBank.com/}}). Machine learning methods are also employed to predict interactions between drugs. The Knowledge4COVID-19 framework is publicly available as a resource in GitHub\footnote{\label{github}\url{https://github.com/SDM-TIB/Knowledge4COVID-19}} and Zenodo\footnote{\url{https://zenodo.org/record/4702125\#.YH4ACu8zaV4}}. Additionally, diverse services are offered to access and explore the KG (e.g., an API\footnote{\label{API}\url{https://github.com/SDM-TIB/Knowledge4COVID-19/tree/main/Exploration-API}} and a public SPARQL endpoint\footnote{\label{KGendpoint}\url{https://labs.tib.eu/sdm/covid19kg/sparql}}). The detailed instructions to access are provided at the project repository\footnote{\url{https://github.com/SDM-TIB/Knowledge4COVID-19/wiki}}. The Knowledge4COVID-19 KG can be created locally following the guidelines\footnote{\url{https://github.com/SDM-TIB/Knowledge4COVID-19/wiki/Running-Knowledge4COVID-19-KG-locally}}.

In summary, the scientific contributions of this work are as follows:
\begin{itemize}
\item 
A novel infrastructure to transform heterogeneous data sources into a knowledge graph based on a unified schema.  The implementation of this infrastructure provides a software pipeline that includes Named Entity Recognition and Named Entity Linking methods, as well as novel mapping rules for aggregating various data retrieved under a unified KG. The resulting KG can be traversed following referenceable resources or queried using SPARQL endpoints or a federated query engine.
\item 
A publicly available KG resource related to COVID-19 integrating information from Scientific Open Data and Publications. This is a product of the aforementioned infrastructure, and allows for the exploration of various sources and data.
\item A deductive system to discover drug-drug interactions in a COVID-19 treatment. This system is built on top of fine-grained representation of Pharmacokinetics drug-drug interactions extracted from scientific open data sources (e.g., DrugBank). 
\item 
A machine learning based drug-drug interaction prediction method, identifying non-documented interactions for treatments related to a specific disease. This produces the predicted COVID-19 related drug-drug interactions that are included in the Knowledge4COVID-19 Data Ecosystem.
\item 
An analysis of the effectiveness and toxicity of COVID-19 treatments, providing drug-drug interactions deduced from the Knowledge4COVID-19 KG and adverse effects of these interactions.
\end{itemize}

\noindent This paper is structured in eight additional sections. \autoref{sec:back} reports the worldwide statistics that summarize the infection situation and presents an overview of the preliminaries. \autoref{sec:approach} defines Knowledge4COVID-19 as a data ecosystem and \autoref{sec:kgcreation} presents the process of knowledge graph creation from the declarative definition using RML mapping rules. \autoref{sec:exploration} describes the Web APIs that enable the traversal of the Knowledge4COVID-19 KG, and the results of the empirical evaluations are reported in \autoref{sec:eval}. The state of the art is summarized in \autoref{sec:rw} and \autoref{sec:resource} describes Knowledge4COVID-19 as a resource. Finally, \autoref{sec:conclusion} wraps up and outlines future work. 
\begin{figure}[ht!]
\centering  
 \includegraphics[width=1.0\columnwidth]{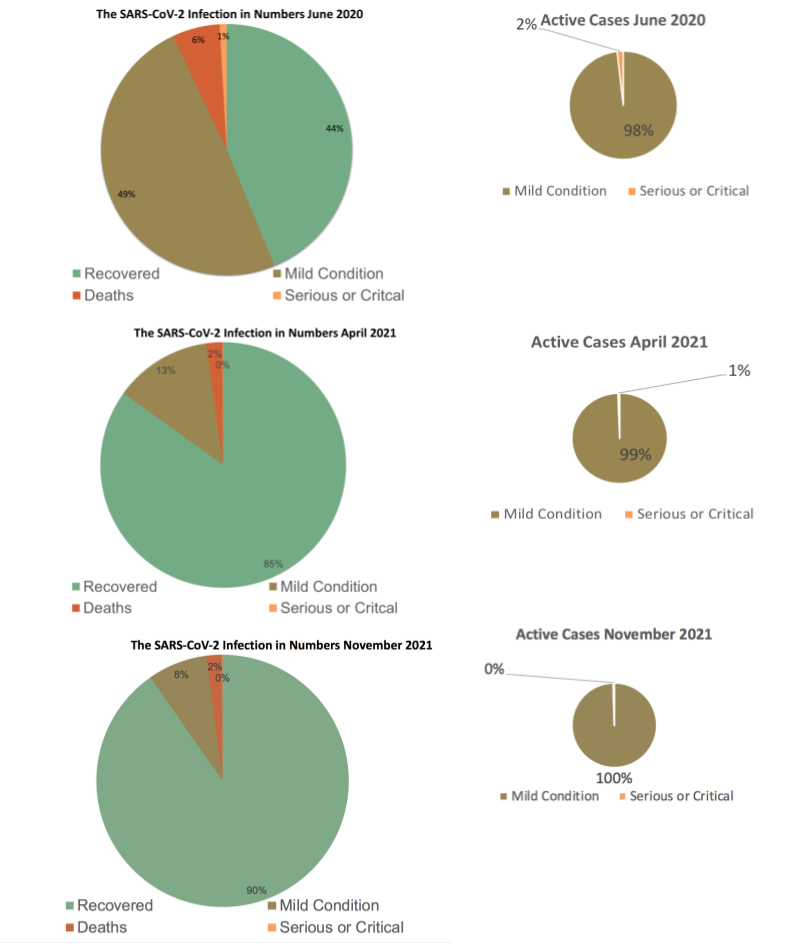}
 \caption{\textbf{SARS-CoV-2 Infections}. Comparison of the severity of infections in June 2020, April 2021, and November 2021. Although the percentages of recovered cases increases significantly, the percentage of new deaths still remains above 2\%.}
  \label{fig:motivatingExample}
\end{figure}
\section{Context and Preliminaries}
\label{sec:back}
\autoref{fig:motivatingExample} depicts world statistics available at Worldometer\footnote{\url{https://www.worldometers.info/coronavirus/}}; numbers of infections by June 2020, April 2021, and November 2021 are summarized. In all three snapshots, at least 98\% of the active infections reported develop mild symptoms, and at most 2\% can be in serious or critical conditions. As can also be observed in the latter, nearly 97\% of the patients who have suffered from COVID-19 have been either categorized as those with a mild condition or have already recovered.

Worldometers\footnote{Data from November 28th, 2021.} also reports weekly new cases concerning the last week of November 2021. The perspective is different when analyzing these reports. The number of infections has increased to 16\% versus 12\% of weekly recovered. Also, the number of new deaths increases by 8\%. The high mortality rate and new cases of infections indicate the unexpected spreading of the virus, still lack knowledge on the infection behavior of populations. Despite the intensity of statistical analyses and related research efforts dedicated to studying the outcome of these infections in certain countries, COVID-19 progression is still unpredictable for most patients, while being many times abrupt for the ones with a severe or critical condition. 

\noindent According to World Health Organization (WHO) statistics\footnote{\url{https://globalhealth5050.org/covid19/age-and-sex-data/}} a broad spectrum of demographic, clinical, and molecular conditions appear to affect the evolution of the disease. 
Although age and sex seem not to be associated with the infection rate, once infected, the mortality rate in men is much higher than in women. Moreover, significant percentages of deaths represent patients above certain ages, as could be expected. Lifestyle variables such as smoking habits also play an essential role. Although regular smokers occur to be significantly underrepresented among those requiring hospital treatment for the illness, smoking emerges to be associated with rapid progression and increased mortality rates. Another factor that seriously affects the fatality rate for COVID-19 seems to be comorbidities, such as cardiovascular diseases, cancer, hypertension, etc. In particular, 80\% of deaths are related to patients with at least one comorbidity, while COVID-19 patients suffering a serious disease (e.g., cancer) seem to develop more rapid progression and appear an increased mortality rate, in contrast to those with no pre-existing chronic medical conditions. Furthermore, the WHO guidelines\footnote{\url{https://www.who.int/publications-detail}} urge clinicians for careful consideration of adverse effects of medications that may be used in the context of COVID-19 and encourage medications that carry the least risk possible of drug-drug interactions with other medicines that a patient with specific comorbidities may be receiving. Researchers should address this need by detecting the risk of documented or even unknown interactions related to specific comorbidities and medications, though looking into big data and identifying relevant patterns.

\subsection{Basic Concepts}
\label{sec:pre}
\textbf{Data Ecosystems}
Data ecosystems (DEs) are data-driven infrastructures that allow different stakeholders to exchange data~\cite{capiello_et_al:DR:2020:11845}.
DEs are furnished with various computational methods to solve interoperability and integrate data while preserving data privacy, security, and sovereignty. DEs can be centralized, and one single node maintains all the data sources shared by the providers. The node also hosts all the services implemented on top of the DE data sources. Contrary, whenever data cannot be moved to a single node and data privacy regulations hinder the materialized and complete data integration of the DE data sources, DEs will be decentralized, i.e., they will be composed of several nodes. Each DE node will be able to perform services and share data management and analytical results.
Semantic data models or ontologies provide the meaning of the data sources in a DE.
Moreover, mapping rules relating to how data sources are defined in terms of the semantic data models are included. 
\\
\noindent
\textbf{Knowledge Graphs (KG):}
 Knowledge graphs~\cite{Hogan2021} are data structures that represent factual knowledge as entities and their relationships using a graph data model. Metadata is part of the KG, as well as taxonomies of entities, relationships, and classes. A KG contributes to the development of a common understanding of the meaning of entities in a domain and provides a formal specification of the properties of these entities. 
 A KG $\mathcal{G}$ can be defined as a data integration system $DIS_\mathcal{G}=\langle O,S,M \rangle$ where $O$ corresponds to the unified schema, $S$ is a set of data sources, and $M$ corresponds to mapping assertions defining concepts in $O$ as conjunctive queries over sources in $S$. The instances of $\mathcal{G}$ are the result of the execution of the $M$ rules over the data sources in $S$. 
 \\
\noindent
\textbf{RDF Mapping Language - RML:}
The RDF Mapping Language (RML)~\cite{DimouSCVMW14} extends the W3C-standard mapping language R2RML to manage heterogeneous data sources represented in various formats, e.g., CSV, XML, JSON, and relational tables. 
These rules, named as \emph{RML triples maps}, define the instances of RDF classes and their properties in terms of a logical source. Attributes from the logical data source of a triples map describe the resources of the corresponding class. 
RML is an RDF triple-oriented mapping language, where a triples map comprises mapping assertions~\cite{namici2018comparing} that define the instances of a class (a.k.a. subjectMap), and the property and object (a.k.a. predicateObjectMap) of the RDF triples where these instances participate as a subject. RML triples maps are expressed in RDF. This allows the exploration and tracing of the definition of the process of KG creation. 
\begin{figure}[ht!]
\centering 
\hspace{0pt}{
 \includegraphics[width=1.0\columnwidth]{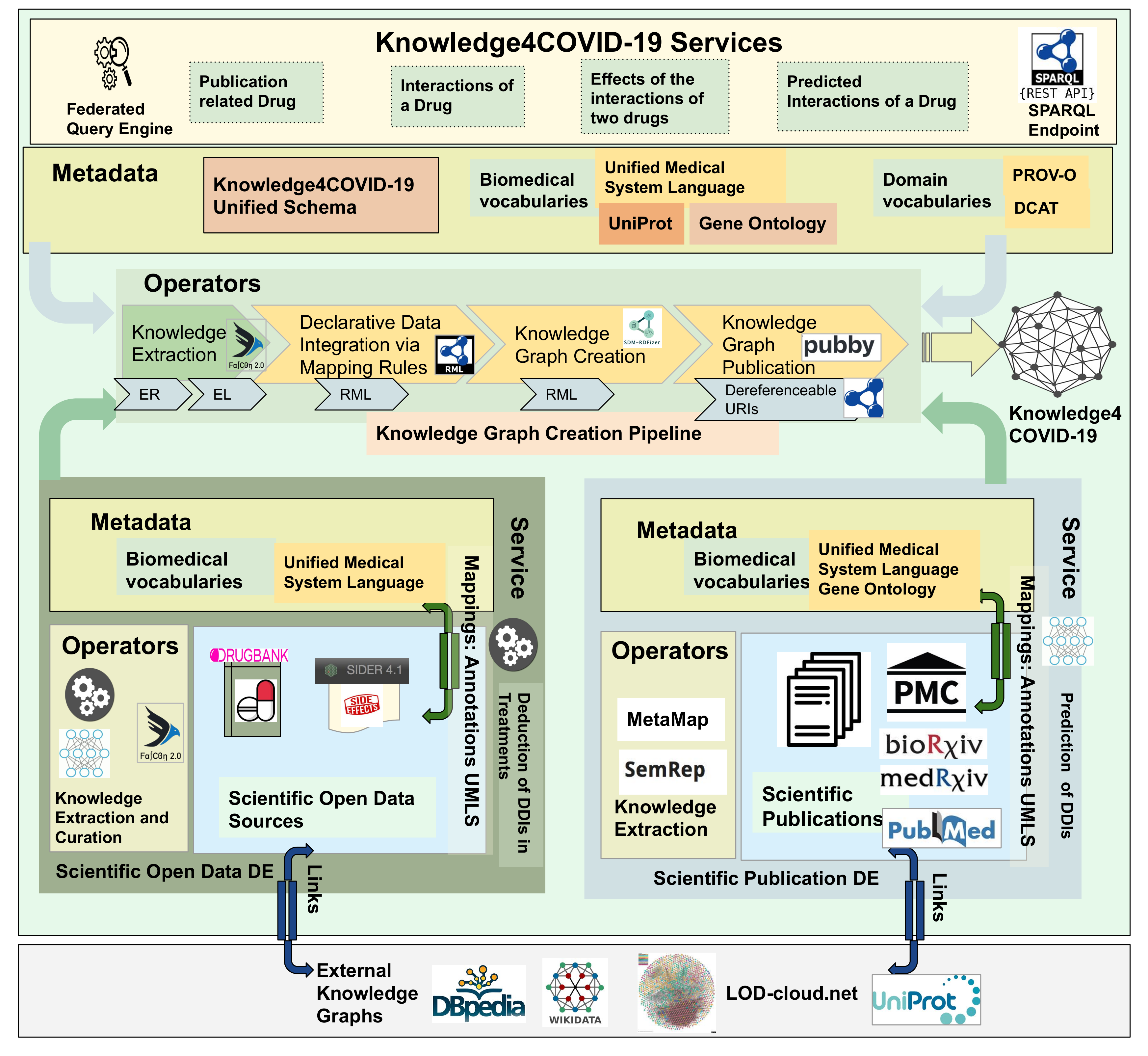}}
 \caption{\textbf{Knowledge4COVID-19 as a Data Ecosystem (DE)}. A Nested Data Ecosystem comprising the Scientific Open Data and Publication DEs. Each DE processes comprises data sources, metadata, and operators to annotate their respective data sources. }
  \label{fig:K4CDE}
\end{figure}
\section{The Knowledge4COVID-19 Data Ecosystem}
\label{sec:approach} 
The Knowledge4COVID-19 framework is a data ecosystem~\cite{oliveira2019investigations}.
A data ecosystem is defined as a 4-tuple \textit{DE=$\langle$Data Sources, Data Operators, Metadata, Mappings, Services$\rangle$}\cite{abs-2105-09312}. 
Data sources represent the collections from where data and knowledge are retrieved. 
 Data operators correspond to functions used for data management (e.g., entity recognition and linking). The metadata component facilitates the specification of the meaning of the data collected from the data sources and the annotation with controlled vocabularies; it also comprises a unified schema that provides an integrated view of the data sources. 
 The mappings align the data sources with the unified schema and describe their meaning.
 Lastly, services exploit the knowledge encoded in the metadata and data operators to satisfy user requirements.
Services include federated query processing, interactions between the drugs of a treatment, predicted drug interactions, or mapping generation. 
 
 \autoref{fig:K4CDE} depicts the components of the Knowledge4COVID-19 DE; it comprises two data ecosystems, one for scientific publications and another for scientific open data. The scientific publication DE includes COVID-19 related literature from PubMed\footnote{\url{https://pubmed.ncbi.nlm.nih.gov/}}, bioRxiv\footnote{\url{https://www.biorxiv.org/}}, medRxiv\footnote{\url{https://www.medrxiv.org/}}, and PubMed Central (PMC)\footnote{\url{https://www.ncbi.nlm.nih.gov/pmc/}}. Scientific open data DE collects data about COVID-19 drugs, their side effects and the adverse events generated by their interactions. The scientific open data DE integrates data extracted from textual descriptions from DrugBank\footref{goDrugBank} and SIDER\footnote{\url{http://sideeffects.embl.de/}}. This section describes these two DEs in detail, while \autoref{sec:kgcreation} describes the pipeline that creates the Knowledge4COVID-19 knowledge graph from the data and knowledge extracted from these two DEs. 
\subsection{The Scientific Open Data DE} 
This data ecosystem makes available data and knowledge about drugs extracted from open data sources. 
\subsubsection{Data Sources}
The Scientific Open Data DE  
integrates medical concepts extracted from open scientific databases. Albeit structured, these datasets may comprise textual attributes that encode relevant entities and relations. 
For example, the drug-drug interaction between Metformin and Hydroxychloroquine is described like \q{The therapeutic efficacy of Metformin can be increased when used in combination with Hydroxychloroquine.}\footnote{\url{https://go.DrugBank.com/drugs/DB00331}}. 
Additionally, the indication of Hydroxychloroquine is presented like 
``Hydroxychloroquine is indicated for the prophylaxis of malaria where chloroquine resistance is not reported, treatment of uncomplicated malaria (caused by P. falciparum, P. malariae, P. ovale, or P. vivax), chronic discoid lupus erythematosus, systemic lupus erythematosus, acute rheumatoid arthritis, and chronic rheumatoid arthritis.``\footnote{\url{https://go.drugbank.com/drugs/DB01611}}.
These descriptions encode relevant facts that can be read and understood by humans. However, further analysis is required to make them understandable by machines. This DE makes used of data operators for named entity recognition to identify entities that correspond to drug related concepts. \autoref{table:SciOpen_sources} describes the data collected from DrugBank \cite{Drugbank}, SIDER \cite{Sider}, and UMLS \cite{UMLS}.  

DrugBank is a Web-accessible database containing information about drugs and their administration routes, mechanisms, proteins, and interactions. Drug-drug interactions can be Pharmacodynamics and Pharmacokinetics.  A pharmacodynamic drug-drug interaction between drugs A and B indicates that both drugs influence in their effects directly, e.g., \q{The risk or severity of QTc prolongation can be increased when Hydroxychloroquine is combined with Acetophenazine}. On the other hand, if drug A has a pharmacokinetic drug-drug interaction with drug B, A alters the disposition (absorption, distribution, elimination) of B, and ends up in the increase or the decrease of B plasma drug concentrations. For example, Abatacept has a pharmacokinetic drug-drug interaction with Hydroxychloroquine, because  \q{The metabolism of Hydroxychloroquine can be increased when combined with Abatacept.}. The Scientific Open Data DE
has collected 769,352	
 and 503,700 Pharmacokinetic and Pharmacodynamics DDIs, respectively. Moreover, 2,421 drug indications and 1,532 toxicities have been collected and processed from DrugBank. 
SIDER is also a Web-accessible database which makes available mechanisms of actions of drugs and their possible adverse effects; 58,945 side effects are collected. UMLS is a controlled vocabulary that comprises terminology, classification, and semantic types and groups of biomedical concepts; 4,536,579 terms are collected together with their definitions, and semantic types and groups. Lastly, following the method proposed by Sridhar et al. \cite{SridharFG16}, two data sources with pairs of drugs that shared at least one protein are computed. CRD are drugs from DrugBank that target at least one protein of the family CYP, while the NCRD drugs also target at least one protein, but it is not of the family CYP.  

\begin{table*}[h!]
\caption{Data sources for the Scientific Open Data DE}
\begin{tabular}{|l|l|l|}
\hline
\textbf{Data Source} &\textbf{Data Type} & \textbf{\#Instances} \\ \hline
DrugBank & Pharmacokinetic DDIs &769,352 \\ \cline{2-3}
\multirow{3}{*} {2022-01-04} 
& Pharmacodynamics DDIs & 503,700 \\ \cline{2-3}
& Drug Indications &2,421 \\ \cline{2-3}
& Drug Toxicities &1,533 \\ \hline
SIDER 2021 & Drug Side Effects &58,945\\ \hline
UMLS Nov 2021 & Medical Concepts &4,864,162\\ \hline
CRD & Pair of drugs that target a CYP protein \cite{SridharFG16} & 345,116\\ \hline
NCRD & Pair of drugs that target a No CYP protein \cite{SridharFG16} &  5,513\\ \hline
\end{tabular}
\label{table:SciOpen_sources}
\end{table*}
\subsubsection{Data Operators}
\begin{figure}[ht!]
\centering
\hspace{0pt}{
     \includegraphics[width=1.03\textwidth, trim = {0 6cm 0 1cm}]{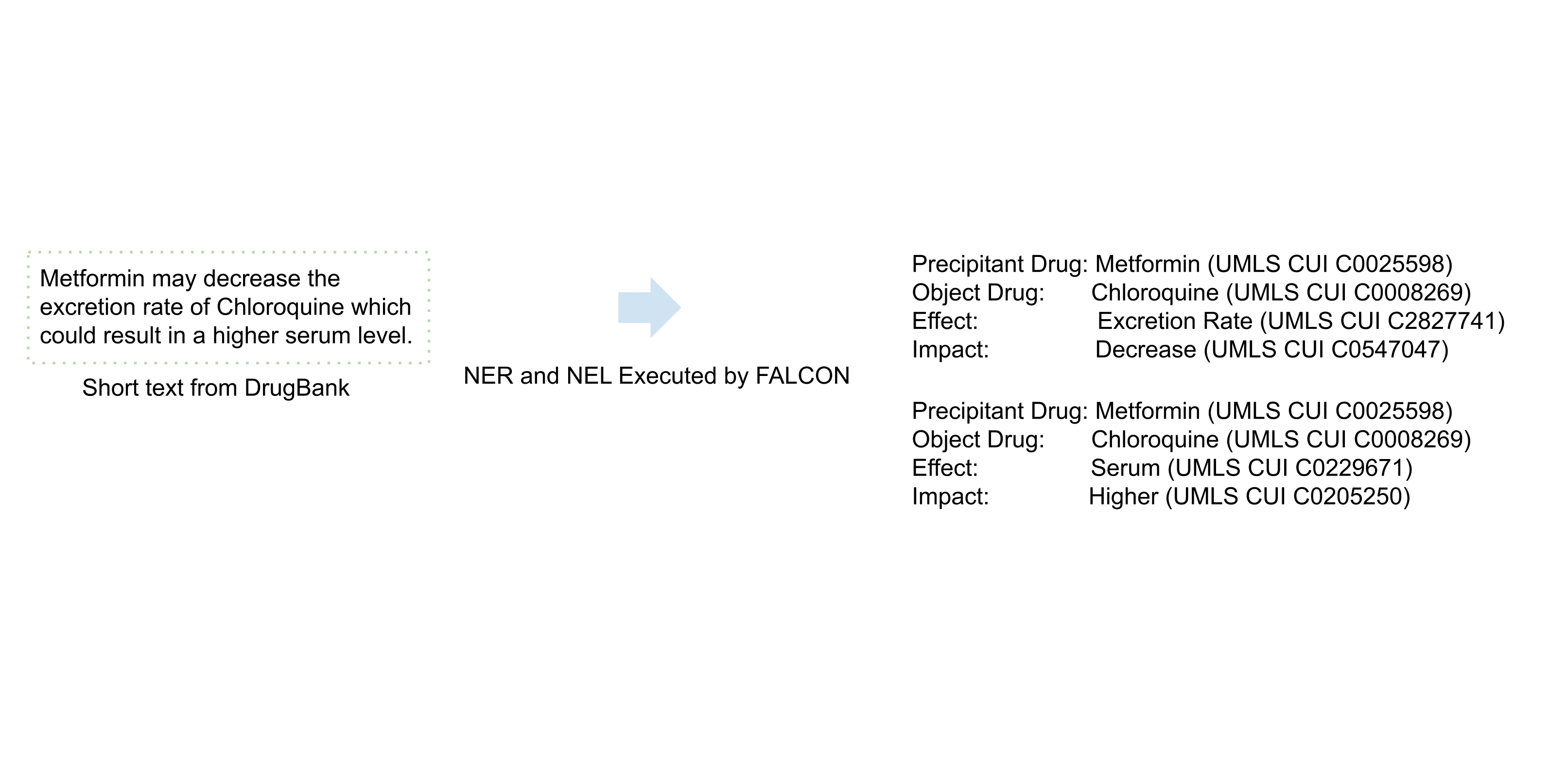}
    }
     \caption{FALCON Recognizes Relevant Entities and Predicates. As a result, a Fine-Grained Representation of Drug-Drug Interactions is part of the Knowledge4COVID-19 KG} 
%  \label{~}
 \label{fig:ddiFALCON}
\end{figure}

The data operators enable the recognition of entities corresponding to drugs, their side effects, and the adverse events caused by their interactions. FALCON ~\cite{sakor2020falcon} recognizes the words corresponding to the drugs that interact and the effect and impact of these interactions. Additionally, the extracted words are linked to terms in UMLS.  As illustrated in Figure \ref{fig:ddiFALCON}, ``Metformin'' and ``Chloroquine'' correspond to the extracted entities from the short text collected from DrugBank. At the same time, ``excretion rate'' and ``decrease'' represent, respectively, the effect and impact of the interaction of ``Metformin'' and ``Chloroquine''. The UMLS identifiers C0025598 and C0020336 are linked to ``Metformin'' and ``Hydroxychloroquine'', while C2827741 and C0547047 are related to ``excretion rate'' and ``decrease'', respectively. FALCON also connects ``Metformin'' and ``Chloroquine''  to their corresponding resources in DBpedia and Wikidata. 

FALCON~\cite{sakor2019old} is also used to extract the Drug-Drug Interactions (DDIs) reported in DrugBank as short texts. We customize FALCON for analyzing the DDI text. Since the DDI text is related to the medical domain, UMLS is utilized as the background knowledge for FALCON. In this case, in addition to recognizing words that correspond to two drugs that interact, FALCON identifies the effect and impact of an interaction. FALCON resorts to the catalog of rules for extracting the previously mentioned types of entities; additionally, a background knowledge base is utilized to determine the semantic type of the extracted entities. Since most of the descriptions of the interactions share similar patterns, i.e., the structure of the sentences is very repetitive, only few extra rules are required to be added to the catalog of rules. The rules were created by replacing each drug mention with a variable (DrugX, DrugY). Out of 1,273,052 drug-drug interactions collected from DrugBank, 320 patterns were recognized; \autoref{table:ddi_patterns} shows a sample of the extracted patterns.

\begin{table*}[h!]
\caption{Overview of extracted DDI patterns. Drugs mentions are in bold. Effect is in Italic. Impact is underlined}
\begin{adjustbox}{width=\textwidth}
\begin{tabular}{|l|}
\hline
\textbf{DDI Patterns}                                                                                                                       \\ \hline
\textbf{DrugY} may \underline{increase} the \textit{anticoagulant activities} of \textbf{DrugX}.                                                                             \\ \hline
the \textit{risk or severity of bleeding and hemorrhage} can be \underline{increased} when \textbf{DrugX} is combined with \textbf{DrugY}.                                   \\ \hline
the \textit{risk or severity of gastrointestinal bleeding} can be \underline{increased} when \textbf{DrugX} is combined with \textbf{DrugY}.                                 \\ \hline
the \textit{risk or severity of bleeding} can be \underline{increased} when \textbf{DrugY} is combined with \textbf{DrugX}.                                                  \\ \hline
the \textit{metabolism} of \textbf{DrugY} can be \underline{decreased} when combined with \textbf{DrugX}.                                                                              \\ \hline
\textbf{DrugX} may \underline{decrease} the \textit{vasoconstricting activities} of \textbf{DrugY}.                                                                          \\ \hline
\textbf{DrugX} may \underline{decrease} the \textit{excretion rate} of \textbf{DrugY} which could result in a \underline{higher} serum level.                                       \\ \hline
\textbf{DrugY} may \underline{increase} the \textit{constipating activities} of \textbf{DrugX}.                                                                              \\ \hline
the \textit{risk or severity of gastrointestinal bleeding and gastrointestinal ulceration} can be \underline{increased} when \textbf{DrugX} is combined with \textbf{DrugY}. \\ \hline
\end{tabular}
\end{adjustbox}
\label{table:ddi_patterns}
\end{table*}

As a result of the knowledge extraction process executed by FALCON, the Scientific Open Data DE makes available fine-grained representation of DDIs. This representation enables the deduction of new drug-drug interactions implemented as a service of this DE. Moreover, these descriptions are also used to validate the prediction tasks implemented in the Scientific Publications DE. 

\begin{table*}[h!]
\caption{Summary of Datalog Predicates. Extensional predicates are ddi(A,E,I,B), member(A,T), treatment(T), rule1(E,I), and rule2(E,I). Intensional predicates are ddi(A,E,I,B,T), toxicity(A,increase,B,T), and effectiveness(A,decrease,B,T).  }\label{tab:notations}
\centering
\footnotesize
\begin{tabular}{|m{4cm}|m{7.2cm}|}
\hline
\textbf{Predicate} & \textbf{Explanation} \\ \hline
ddi(A,E,I,B) & Pharmacokinetic drug-drug interaction between $A$ and $B$. Precipitant drug $A$ generates effect $E$ (e.g., absorption, excretion, metabolism, serum concentration) with impact $I$ (e.g., increase or decrease) in object drug $B$.\\ \hline
ddi(A,E,I,B,T) & Pharmacokinetic drug-drug interaction between $A$ and $B$ in treatment $T$. Precipitant drug $A$ generates effect $E$ (e.g., absorption, excretion, metabolism, serum concentration) with impact $I$ (e.g., increase or decrease) in object drug $B$.\\ \hline
rule1(E,I) & Combinations of effect $E$ with impact $I$ that alter the toxicity of an object drug. \\ \hline
rule2(E,I) & Combinations of effect $E$ with impact $I$ that alter the effectiveness of an object drug. \\ \hline
treatment(T) & $T$ is a medical treatment\\ \hline
member(A,T) & $A$ is a drug in the medical treatment $T$\\ \hline
toxicity(A,increase,B,T) & The precipitant drug $A$ increases the toxicity of object drug $B$ in treatment $T$\\ \hline
effectiveness(A,decrease,B,T) & The precipitant drug $A$ reduces the effectiveness of object drug $B$ in treatment $T$\\ 
\hline
\end{tabular}
\end{table*}

\subsubsection{Data Services} 
The Scientific Open Data DE implements a deductive system that enables to deduce drug-drug interactions among a multi-drug treatment whose interactions may reduce the effectiveness of the treatment or increase the number of toxicities. 
The deductive system is defined in terms of Datalog rules; it exploits the fine-grained representation of the DDIs interactions generated by FALCON. 
The execution of this deductive system is grounded on the results of deductive databases \cite{CeriGT89} to compute the minimal model that includes the instances of the deduced drug-drug interactions in a treatment. 
The minimal model corresponds to the
fixed-point of the assignments of the values of variables in the deductive system rules. Since rules free of negations compose the deductive system, the minimal model is computed in polynomial time in the size of the number of treatments and drug-drug interactions generated by FALCON.
The approach proposed by Rivas and Vidal~\cite{DBLP:conf/kcap/RivasV21} is followed to implement this data service. 
The extensional database corresponds to statements about interactions between drugs extracted by FALCON. On the other hand, the intensional database comprises a set of Horn clauses that define the conditions to be met by the drugs whose interactions may reduce the effectiveness of a treatment or increase the number of toxicities. 
This intensional database relies on the fact that pharmacokinetic drug-drug interactions cause that the concentration of one of the interacting drugs (a.k.a. object) is altered when it is combined with the other drug (a.k.a. precipitant). Thus, the rate of absorption, distribution, metabolism, or excretion of the object drug is affected. 
Whenever the object drug absorption is decreased (resp. increased) the bioavailability of the drug is also affected.
Furthermore, any alteration in the metabolism or excretion of the object drug has consequences on the therapeutic efficacy and toxicity of the drug. The following Datalog rules state the effect of pharmacokinetic DDIs. 
Considering the predicates in \autoref{tab:notations}, the intensional database defines the toxicity effects of drug-drug interactions in a treatment: 
\begin{align*}
ddi(A,E,I,B),treatment(T),member(A,T),member(B,T)\rightarrow \\ 
ddi(A,E,I,B,T). \\
ddi(A,E,I,B,T),rule1(E,I)\rightarrow \\ 
toxicity(A,increase,B,T). \\
toxicity(A,increase,B,T),toxicity(B,increase,C,T)\rightarrow\\ toxicity(A,increase,C,T). \\
toxicity(A,increase,B,T), ddi(B,E,I,C,T)\rightarrow ddi(A,E,I,C,T).\\ 
\end{align*} 

The conditions to reduce effectiveness are defined as follows:
\begin{align*} 
ddi(A,E,I,B,T),rule2(E,I)\rightarrow \\
\textit{effectiveness}(A,decrease,B,T). \\
\textit{effectiveness}(A,decrease,B,T),\textit{effectiveness}(B,decrease,C,T)\rightarrow \\ \textit{effectiveness}(A,decrease,C,T).
\end{align*} 
The extensional database includes the following ground predicates:
\begin{align*}
rule1(serum, increase). \\
rule1(metabolism, decrease). \\
rule1(absorption, increase). \\
rule1(excretion, decrease).\\
rule2(serum, decrease). \\
rule2(metabolism, increase). \\
rule2(absorption, decrease). \\
rule2(excretion, increase).\\
\end{align*}
Additionally, a graph traversal method is implemented to compute the drugs that affect the most the effectiveness or toxicity of a treatment drug. The implemented method creates a directed graph from drug-drug interactions with the extensional facts and deduced of the Datalog rules. 
The direction of an edge from node A to B denotes that A is the precipitant and B is the object of the interaction. 
Drugs that affect the most the effectiveness or toxicity of a treatment drug are defined in terms of the middle-vertices in the wedges~\cite{DBLP:conf/kcap/RivasV21}, or paths with two directed edges~\cite{10.14778/3447689.3447702}, in the directed graph that represents the drug-drug interactions among the drugs of a treatment. The middle-vertex of a wedge is both the object drug of one interaction, and the precipitant drug of the other interaction. Thus, drugs that correspond to middle-vertices of $N$ wedges in a treatment $T$, correspond to drugs that cause $2*N$ different drug-drug interactions in that treatment.
\begin{figure*}
    \centering
    {\includegraphics[width=\textwidth]{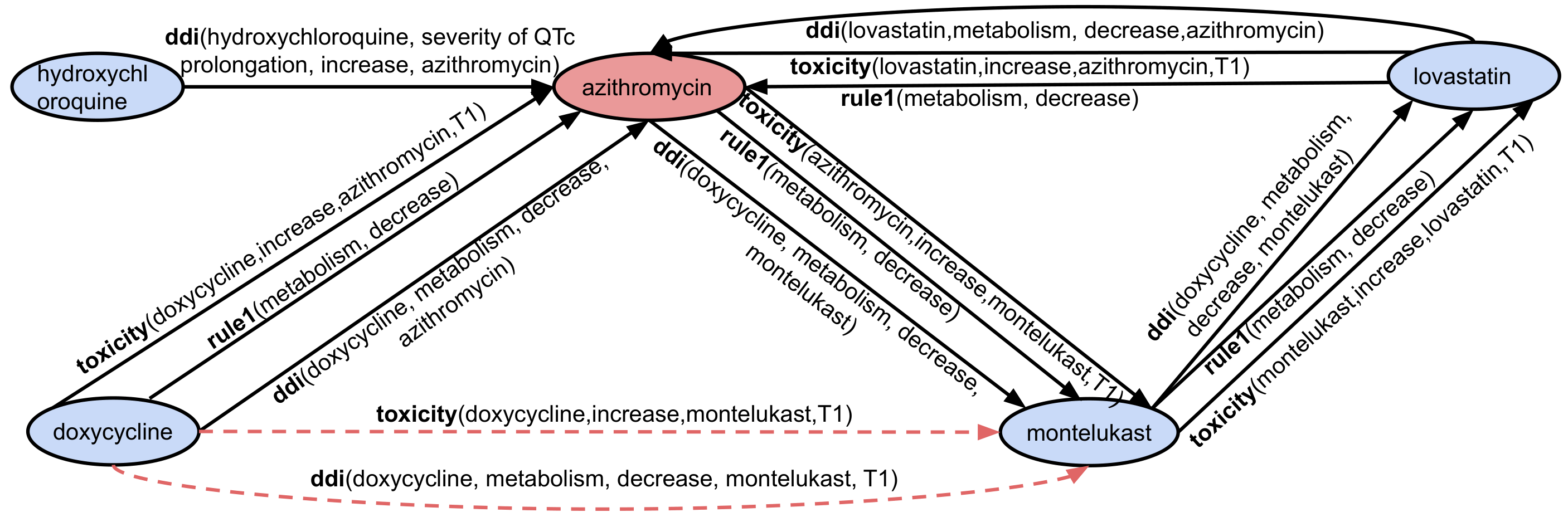}}
    \caption{\textbf{COVID-19 treatment}. Example of deducing DDIs and computing wedges. The red arrows represent the DDIs deduced, and the red node represents the drug with the higher absolute frequency of being the wedges middle-vertex.}
    \label{fig:exampleTreatment}
\end{figure*}

\autoref{fig:exampleTreatment} depicts an exemplar treatment composed of  five drugs. DDIs are represented by the predicates summarized in \autoref{tab:notations}.
By evaluating the Datalog program, a new DDI is deduced and represented in red.
This evaluation deduces that doxycycline decreases the metabolism of montelukast in the treatment T1 and that doxycycline increases toxicity of montelukast.
For the sake of simplicity, a single deduced DDI is depicted, even if the 
Datalog program deduces five new DDIs.
Computing the absolute frequency of a drug being the wedges middle-vertex identifies the drugs that affect the most the effectiveness or toxicity of drug treatment.
In this case, azithromycin is the drug with the higher absolute frequency of being the middle-vertex of the wedges in the graph (i.e., absolute frequency of three); it is followed by montelukast with a value of two and lovastatin with a value of one.
Note that after removing azithromycin from the treatment, there is only one DDI between montelukast and lovastatin, i.e., 83.3\% of the DDIs are eliminated.  

In \autoref{sec:toxicity}, we compare the drug-drug interactions deduced by the previously described deduction system and existing tools that discover drug-drug interactions in a treatment. The results of this evaluation suggest that middle-vertices, with high frequency in the directed graph of a therapy, correspond to the drugs that produce more toxicities. Therefore, identifying frequent middle-vertices in the directed graph that models a treatment provides a computational method for discovering toxic medications in treatment. 

\subsection{The Scientific Publications DE}
\label{sci-pub-de}
\sloppy
The Data Ecosystem of Scientific Publications
comprises the components extracting relevant medical concepts from scientific publications.
\subsubsection{Data Sources}
The Scientific Publications DE collects data from the following data sources: CORD-19 \cite{CORD-19}, PubMed \footnote{\url{https://pubmed.ncbi.nlm.nih.gov/}}, and PubMed Central (PMC)\footnote{\url{https://www.ncbi.nlm.nih.gov/pmc/}}, enriched with information from certain ontologies. 
CORD-19 is a collection of scientific papers about COVID-19 and related coronavirus; the version by 2021-03-01 includes 460,772 publications. PubMed is Web-accessible engine to primary access scientific publications from the MEDLINE database. The Scientific Publications DE has 
and harvested  articles from PubMed and PubMed Central (PMC) until April 2022, including the MeSH topic 'covid-19'. 
\autoref{table:SciPub_sources} describes the full list of data sources and ontologies used by the Scientific Publications DE.

\begin{table*}[h!]
\caption{The full list of data sources and ontologies used for the Scientific Publications DE}
\centering
\begin{tabular}{|l|l|l|l|}
\hline
 \textbf{Sources} &
 \textbf{\#publications} &
\textbf{Ontologies} &
 \textbf{\#annotations} \\ \hline
 PubMed &
 106,150 &
 MeSH &
 1,356,578 \\
 PMC &
 26,105 &
 Gene Ontology &
 125,629 \\
 CORD-19 &
 460,772 &
 Disease Ontology&
 5,129 \\ \hline

\end{tabular}
\label{table:SciPub_sources}
\end{table*}

\subsubsection{Data Operators}
The natural language processing (NLP) tools MetaMap\footnote{\url{https://metamap.nlm.nih.gov/}} and SemRep\footnote{\url{https://semrep.nlm.nih.gov/}} are utilized to recognize drugs and diseases from the titles and abstracts of the integrated articles\footnote{\url{https://github.com/SDM-TIB/Knowledge4COVID-19/wiki/CORD-19-Publication-Processing}}, and also from the full texts of articles that are available in PMC. The Unified Medical Language System (UMLS) is used to describe the extracted medical entities using a controlled vocabulary of medical terms. Moreover, the Medical Subject Headings (MeSH) thesaurus, along with some Open Biological and Biomedical Ontology (OBO) Foundry ontologies, are also harvested in order to retrieve topic annotations and hypernymic relations of drugs and diseases. 

In total, 542,672 publications are annotated with semantics relations from UMLS\footnote{\label{meta3}\url{https://www.nlm.nih.gov/research/umls/META3\_current\_relations.html}} (e.g., ASSOCIATED\_WITH, TREATS, CAUSES), adverse events (e.g., Dyspnea increase, Confusion increase), disorders (e.g., Colorectal cancer, Bladder cancer), phenotypes (e.g., Allergic Reaction, Hemorrhage), and drugs (e.g., Becaplermin, Naloxone). Furthermore, metadata of the processed publications (e.g., title, authors, publication date, journal name, and citation number) describes the main attributes of the scientific publications.

\subsubsection{Data Services} 
Despite the wide adoption of MetaMap and SemRep tools, their effectiveness is far from perfect~\cite{kilicoglu2020broad}. Thus, triples resulting from applying those NLP tools on publications tend to be the most noisy part of the knowledge graph. 
To overcome this quality challenge, we need to apply some kind of error detection mechanism for the Scientific Publications Graph refinement~\cite{yuan2020constructing,zhang2018automatic}. In our case, we have experimented with various approaches, such as graph embeddings~\cite{bordes2013translating}, path ranking solutions (PaTyBRED)~\cite{melo2017detection} and a hybrid approach called PRGE (Path Ranking Guided Embeddings)~\cite{bougiatiotis2021guiding}. PRGE method uses the PaTyBRED path ranking technique, in order to produce confidence scores for all the triples of a graph. It then uses those scores in order to guide the TransE embedding method focusing on the probably correct triples, during the graph embedding creation. This is realised by incorporating triple confidence scores in the embedding Loss function, guiding thus the training
procedure to put less emphasis on noisy triples. The selected approach results in a final confidence score for each triple of the graph in the range of [0-1]. Deciding a cut-off confidence threshold below which all triples will be considered as erroneous provides a trade-off between quality and the amount of data that will be produced. In our case, we selected a median threshold of 0.5, in order to keep the majority of the graph triples. Applying this method to the Scientific Publications DE Graph resulted into a 40\% of the total triples identified as possibly erroneous.

\noindent\textbf{Scientific Publications DE analysis:} As a next step, we apply a predictive analysis on the Scientific Publications DE, in order to identify previously unknown adverse effects of drug combinations, in the form of drug-drug interaction relations. For this purpose, a machine learning method that exploits patterns unveiled from contextual information of the Scientific Publications DE to predict potential drug-drug interactions is implemented. This method is based on the analysis of the Scientific Publications Graph~\cite{nentidis2020iasis} that results from the natural language processing and semantic indexing of biomedical publications and open resources, as described above. The Scientific Publications Graph constitutes an integral part of the Knowledge4COVID-19 KG, representing the structured information extracted from relevant publications in the form of triples. Drugs included in DrugBank are also considered a part of this graph, relating these with specific targets, diseases, and other biomedical entities identified in literature text, through a set of semantic relations from the UMLS Semantic Network\footref{meta3}.

\begin{figure}[ht!]
\centering  
\hspace{0pt}{
     \includegraphics[width=1.0\textwidth]{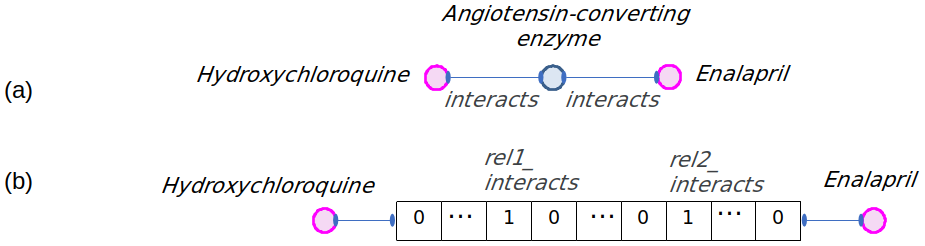}
    }
     \caption{ An example of (a) a semantic path  between drugs Hydroxychloroquine and Enalapril, (b) transformation into a feature vector. } 
%  \label{~}
 \label{fig:ddi}
\end{figure}

\noindent\textbf{Prediction of new DDIs:} The problem of predicting new drug-drug interactions is addressed as a binary classification problem for interacting/non-interacting drug pairs in the Scientific Publications Graph. The result of this classification provides a set of drug pairs with none previously known interaction, marked as False Positives, that our classifier identifies as interacting with a certain confidence score. These predictions can provide an indication of potential interactions to pharmaceutical experts that have not been previously documented. 
To this end, the aforementioned machine learning technique focuses on the analysis of the undirected semantic paths connecting different pairs of drugs in the Scientific Publications Graph. This method is called Drug-Drug Interaction prediction on a Biomedical Literature Knowledge Graph (DDI-BLKG)~\cite{bougiatiotis2020drug}. Each one of these paths includes a sequence of semantic relations of length $n$ that are aggregated into feature vectors representing the frequency of each relation in a specific position $(1,n)$. As an example, if Hydroxychloroquine and the Diabetes-related Enalapril both interact with the target Angiotensin-converting enzyme, this provides the undirected path and the respective feature vector (\autoref{fig:ddi}).

\noindent Let $D$ be the number of relevant drugs examined, where relevance is determined by the existence of such drugs in COVID-19 related publications. Aggregating all possible paths between pairs of drug nodes, we generate a big dataset of $(D-1)!$ feature rows that denote relations' frequency in specific positions, as illustrated above. Each feature row is of size ($n$ x $r$), where r denotes the number of different relation types. In our case, maximum path length is set to 3 $(n=3)$, as this has provided the best trade-off between data size and accuracy. Also, 35 unique relation types are used from the UMLS Semantic Network $(r=35)$. Therefore. (3 x 35) features are calculated for every pair and are used to train a Random Forest classifier that is able to effectively discriminate between two classes: interacting and non-interacting pairs, based on the respective label extracted from a gold dataset.

In order to generate the final set of predictions, the Random Forest classifier is trained using all COVID-19 related pairs (where at least one of the two drugs is mentioned in Drugbank as COVID-19 experimental treatments\footnote{\url{https://go.drugbank.com/covid-19\#drugs}}), denoted as positives in DrugBank. Testing the classifier for all possible remaining COVID-19 related drug pairs, which are not known to be interacting, produces 8,925 unknown drug-drug interaction predictions in total, with a certain confidence score within a range of [0,1]. The critical threshold of this score is considered to be 0.5, meaning that drug pairs with a score $<0.5$ are less possible to be interacting, while pairs with a score $>0.5$ represent the most possible interactions.

\section{The Knowledge4COVID-19 Knowledge Graph}
\label{sec:kgcreation}
This section the Knowledge4COVID-19 DE in terms of the 
pipeline for the creation of Knowledge4COVID-19 knowledge graph (KG), the linking to existing KGs (e.g., DBpedia and Wikidata), and the techniques of federated query processing implemented on top of Knowledge4COVID-19 KG.
The Knowledge4COVID-19 DE relies on annotations from UMLS,  DBpedia, and Wikidata to solve entity alignment. 
%NLP techniques for free text analysis, semantic indexing, and annotation based on the UMLS lexicon, enable the integration of a variety of non-structured data sources. Moreover, entity recognition and linking from textual attributes (e.g., labels and short descriptions) allow for the alignment to DBpedia and Wikidata. 
The execution of 221 RML mapping assertions --manually defined by two knowledge engineers and curated by two more-- transforms the structured representation of the data sources, annotations, and alignments into the Knowledge4COVID-19 KG.

\begin{figure}[t!]
\centering  
\hspace{0pt}{
 \includegraphics[width=1.0\columnwidth]{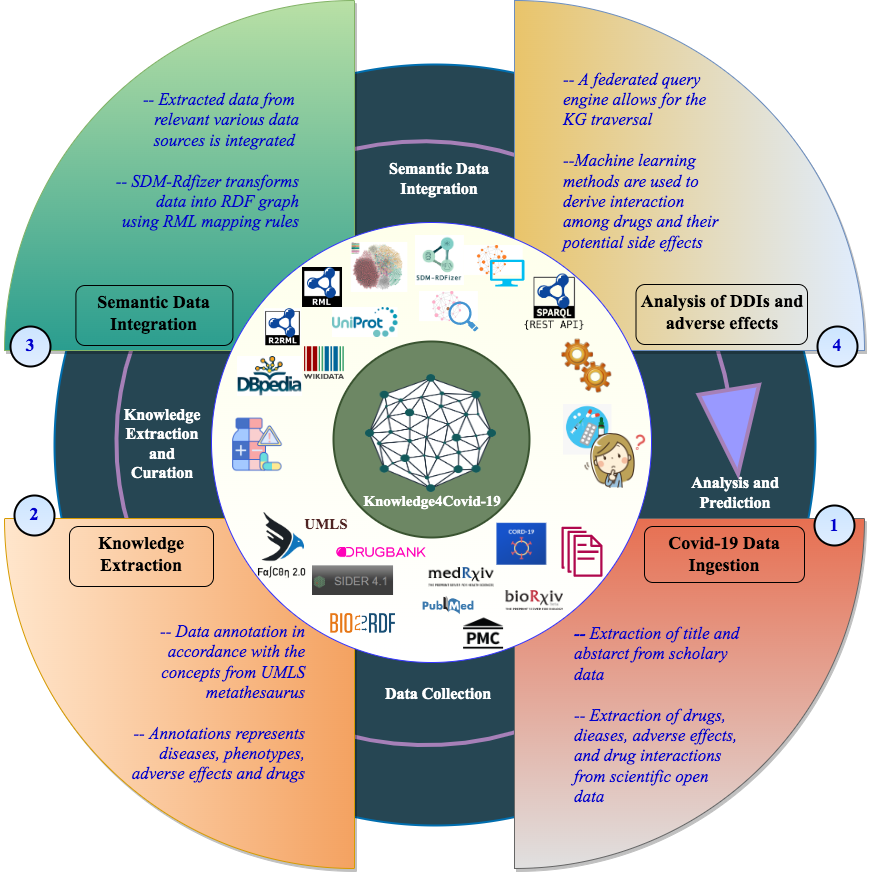}}
 \caption{\textbf{The Knowledge4COVID-19 KG Pipeline}. Steps followed during the transformation of heterogeneous data into the Knowledge4COVID-19 KG. UMLS annotations provide the basis for entity alignment and data integration.}
  \label{fig:K4CKG}
\end{figure}

\autoref{fig:K4CKG} depicts the steps of the KG creation process. Steps 1 and 2 are done at the level of Scientific Open Data and Publications DEs, while steps 3 and 4 are conducted at the level of Knowledge4COVID-19 DE (\autoref{fig:K4CDE}) to create the Knowledge4COVID-19 KG. First, data is ingested and described in terms of metadata (step 1), e.g., title and abstract of the publications, and drug-drug interactions. Knowledge extraction methods recognize biomedical entities from textual data and link them to UMLS, and to resources in DBpedia, Wikidata, Uniprot, and DrugBank. A total of 12,223,409 UMLS annotations have been extracted by FALCON. 
These annotations are used for solving entity alignment and semantic data integration of biomedical entities in the Knowledge4COVID-19 KG (e.g., drugs, phenotypes, side effects, and adverse events). 
Moreover, there are 3,739,445 links to DBpedia, 3,476,435 links to Wikidata, 5,248 links to the Uniprot RDF KG, and 3,427 links to DrugBank. 

The shared data sources are mapped to the Knowledge4COVID-19 unified schema. 
%These definitions are declaratively specified using the RDF Mapping Language (RML).
SDM-RDFizer~\cite{iglesias2020sdm} transforms these shared data into an RDF graph by executing the RML mapping rules. SDM-RDFizer implements optimized data structures that are exploited during the execution of RML mapping rules to speed up the KG creation process~\cite{iglesias2020sdm}. %Experimental results report by Iglesias et al.~\cite{iglesias2020sdm} suggest that these process can reduce execution time in at least one order of magnitude with respect to state-of-the-art RML engines (e.g., RMLMapper\footnote{\url{https://github.com/RMLio/rmlmapper-java}} and RocketRML\cite{csimcsek2019rocketrml}).
The Knowledge4COVID-19 KG is published following the Linked Data principles. A linked data interface using Pubby\footnote{\url{https://github.com/cygri/pubby}} is provided; thus, all the URIs can be dereferenced. Additionally, a SPARQL endpoint allows for querying processing on top of the Knowledge4COVID-19 KG, while the federated query engine, DeTrusty~\cite{DeTrusty}, evaluates SPARQL queries over the federation of 
the Knowledge4COVID-19 KG, DBpedia, Wikidata, and UniProt RDF. Additionally, various API REST services are offered to traverse the Knowledge4COVID-19 KG, and analyze drug-drug interactions and side effects (step 4).

\subsection{The Knowledge4COVID-19  Unified Schema}
The Knowledge4COVID-19 unified schema comprises concepts that provide abstract representations of the entities present in the data sources. 
\begin{figure}[t!]
\centering  
\hspace{0pt}{
      \includegraphics[width=1.0\linewidth]{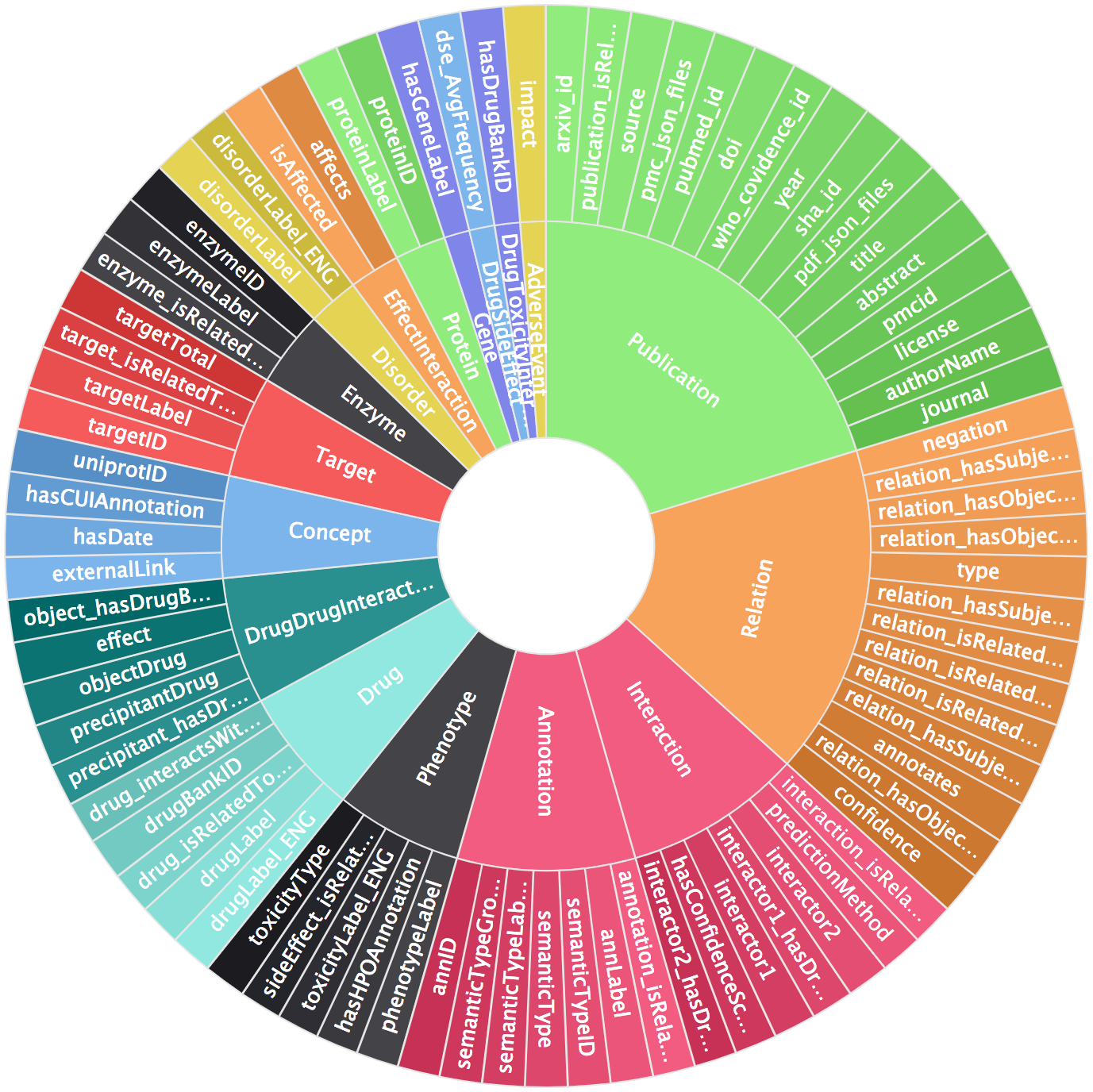}}
     \caption{{\bf The Unified Schema}. Classes and properties.}
     \label{fig:classes}
\end{figure}
Each generic concept of a type or category is defined as a Class in OWL. These concepts represent annotations from controlled vocabularies, drugs, COVID-19 treatments and drugs, disorders, phenotypes, adverse events, enzymes, targets, side effects, scientific publications, and interactions between drugs, drugs and side effects, and drugs and their targets. The current version of the unified schema is composed of 67 classes, 37 object properties, 49 data type properties, and eight annotation properties.
\autoref{fig:classes} shows examples of classes and properties of the Knowledge4COVID-19 unified schema. The inner circle in the \autoref{fig:classes} displays 17 classes of the unified schema; each class is shown in a different color. The outer circle, however, illustrates examples of the properties categorized by the classes. Each group of properties shown in the same color as one class represents all the properties which domains are the same class; in average, a class has in 3.7 properties. Following the Global as View (GAV) modeling approach~\cite{lenzerini2002data}, we define the classes in the unified schema such that they involve all the concepts represented in data sources and recognized by a domain expert. Similarly, the properties are defined considering the domain specific relations between the concepts residing in different data sources.

In defining the unified schema concepts, we exploit two available unified schemas corresponding to two different biomedical knowledge graphs: iASiS\footnote{\url{http://ontology.tib.eu/iasis/}} and BigMedilytics\footnote{\url{http://ontology.tib.eu/bigmedilytics/}}. 
Additionally, the Knowledge4COVID-19 unified schema concepts (i.e., classes and properties) are related via the \texttt{owl:equivalentClass} and \texttt{owl:equivalentProperty} predicates to concepts in DBpedia, Wikidata, Uniprot, the Open Biological and Biomedical Ontology, the Semanticscience Integrated Ontology, and Dublin Core. In total, 17 concepts are mapped to at least one concept in these ontologies. 

\begin{figure*}[t!]
\centering
    \begin{subfigure}{.45\linewidth}
    \centering
        \includegraphics[width=1.0\linewidth]{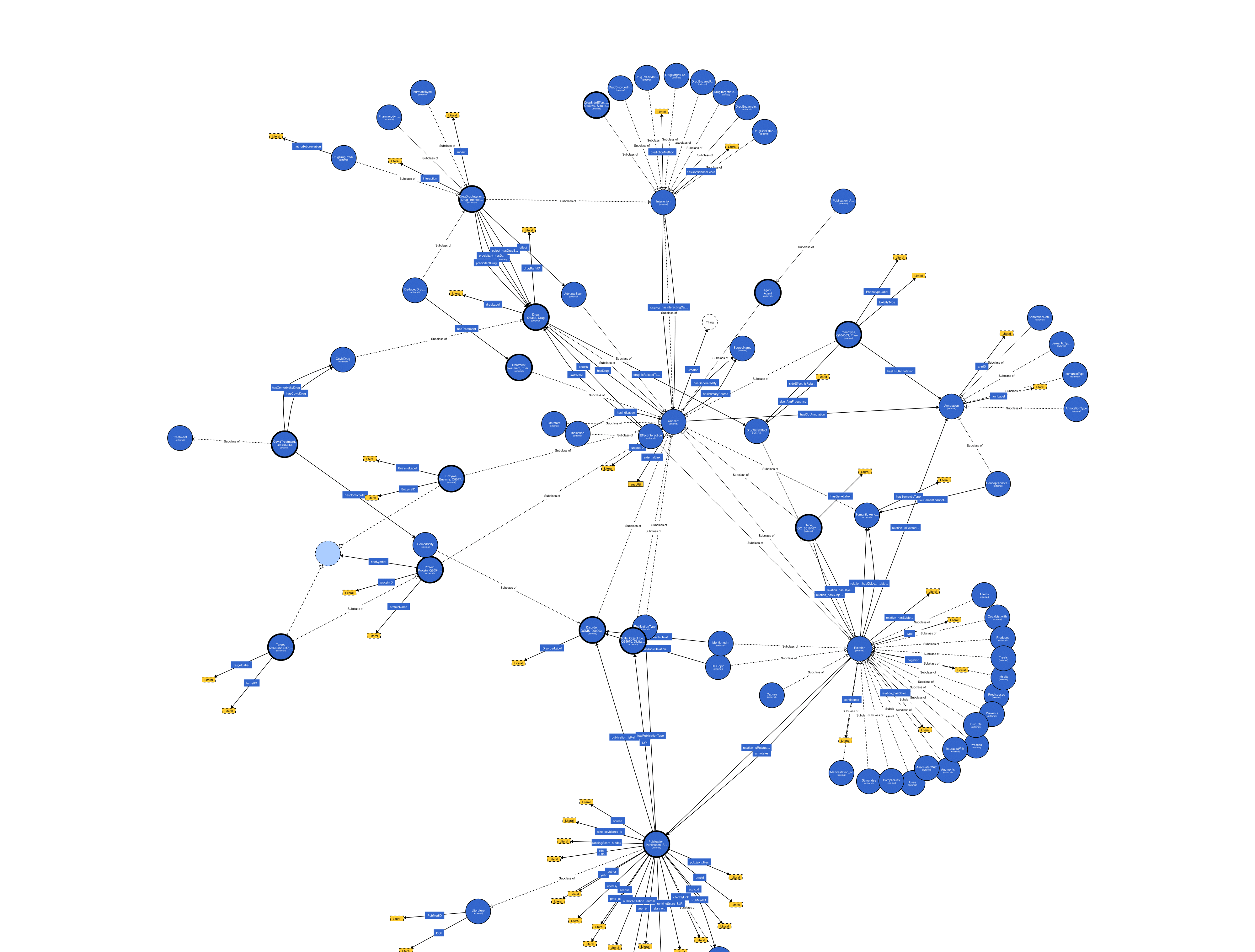}
   \vspace{-0.25cm}
    \caption{Classes and properties}
    \label{fig:visualization}
    \end{subfigure}
     \begin{subfigure}{0.5\linewidth}
    \centering
        \includegraphics[width=1.0\linewidth]{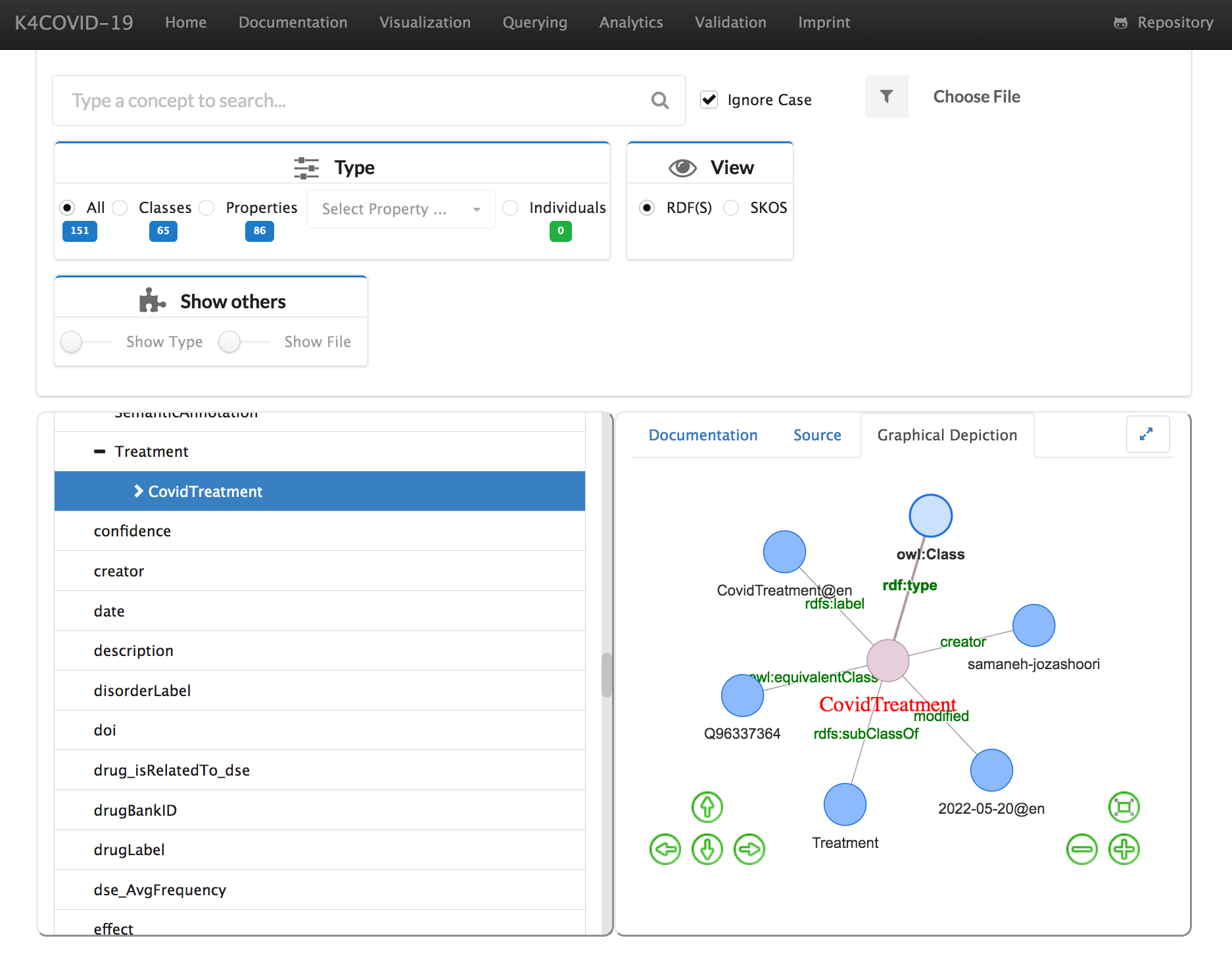}
   \vspace{-0.25cm}
    \caption{Metadata of the class covid-19:CovidTreatment}
      \label{fig:documentation}
    \end{subfigure}
    \caption{{\bf The Knowledge4COVID-19 unified schema}. VoCol Visualization of the classes, and data and object properties.}
    \label{fig:VoCol}
\end{figure*}

The unified schema is publicly available as a VoCol repository supported by TIB \footnote{\url{http://ontology.tib.eu/K4COVID-19/}}. VoCol~\cite{HalilajPG0ACL16} provides a loose coupling of components for validation, querying, analytics, visualization, and documentation on top of a standard Git repository. VoCol also provides an interface for specifying queries against the unified schema and ontology management features that enable the visualization and exploration of the ontology.  Finally, the documentation describing the metadata of each class and property can be consulted, as well as basis analysis describing the number of classes and properties that comprise the unified schema.  \autoref{fig:visualization} depicts the Knowledge4COVID-19 unified schema visualized by VoCol. The metadata describing each of the depicted concepts can be accessed at VoCol \footnote{\url{http://ontology.tib.eu/K4COVID-19/documentation}}. \autoref{fig:documentation} presents the description of the class \texttt{covid-19:CovidTreatment}, which groups of COVID-19 drugs and the drugs of common comorbidities.

\subsection{Mapping the Data Sources into the Unified Schema}

\begin{figure}[b!]
\centering  
\hspace{0pt}{
     \includegraphics[width=1.0\linewidth]{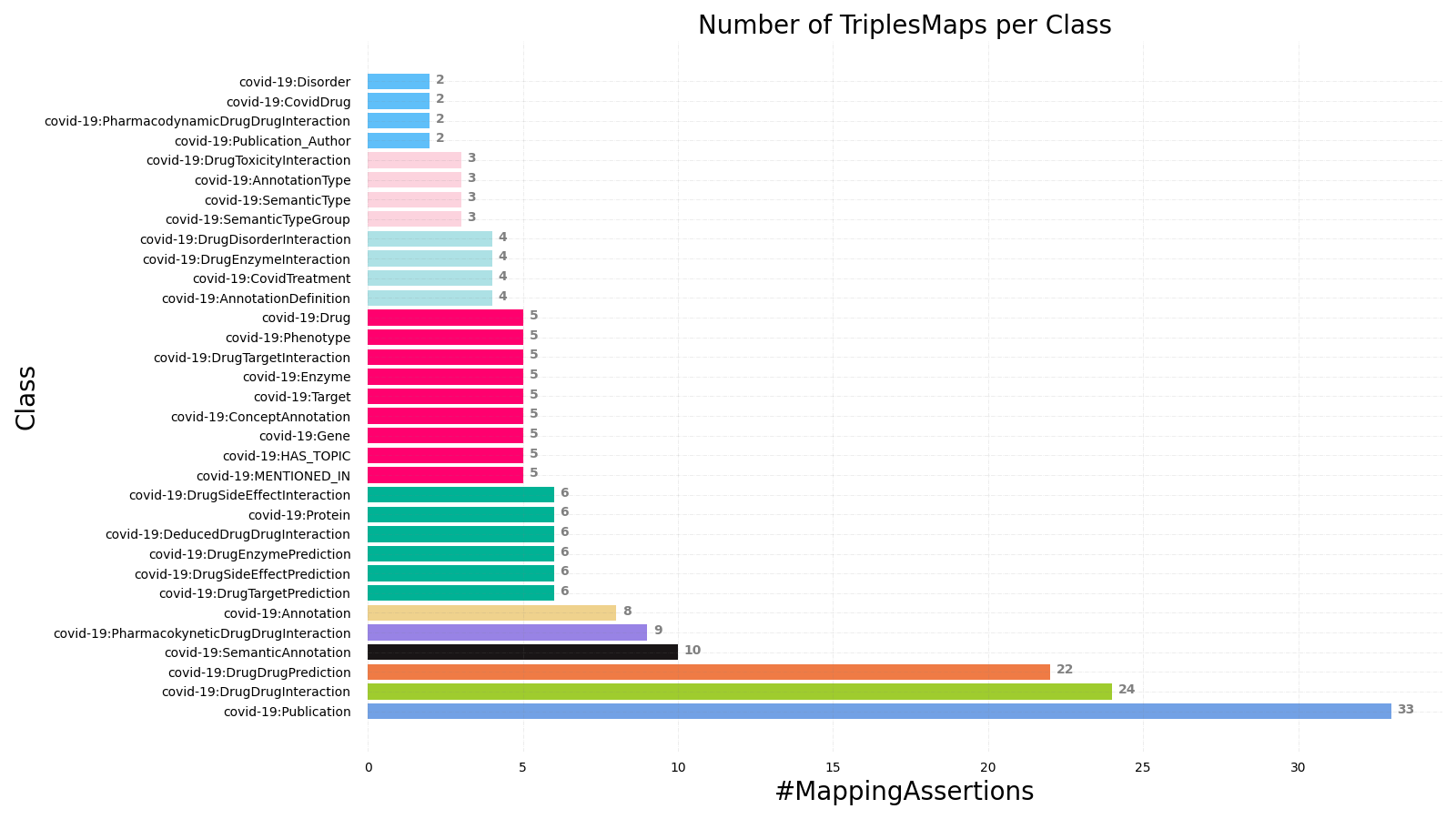}}
     \caption{{\bf Mapping Assertions in RML Triples Maps}. Number of Mapping Assertions (i.e., subject and object maps) per classes.}
      \label{fig:MotGraphPRECIS}
\end{figure}

Classes and properties in the unified schema are defined in terms of the attributes in the data sources by means of RML triples maps.
The Knowledge4COVID-19 KG is defined using 57 RML triples maps that comprise 223 mapping assertions (i.e., subject or object Map). 
\autoref{fig:MotGraphPRECIS} presents the number of mapping assertions of the RML triples maps that define each of the unified schema classes and their properties. For example, the class \texttt{COVID-19:DrugDrugInteractionPrediction} is defined using 22 mapping assertions, and   \texttt{COVID-19:Publication} is the class with the greater number of properties and is defined by 33 mapping assertions. A SPARQL endpoint with the unified schema and the triples maps is publicly available\footnote{\url{https://labs.tib.eu/sdm/covid19kg-mappings/sparql}}. 

\autoref{fig:query1} presents a SPARQL query that collects the information about the mapping rules that define the class \texttt{COVID-19:Publication}. The results of this query evaluation include the data source from where the data is collected, and per predicate of the class, the attribute(s) of the corresponding data source used to populate the predicate.

\begin{figure}[t]
    \begin{lstlisting}[style=sparql,basicstyle=\footnotesize]
PREFIX rr:  <http://www.w3.org/ns/r2rml#>
PREFIX rml: <http://semweb.mmlab.be/ns/rml#>
PREFIX COVID-19:  <http://research.tib.eu/covid-19/vocab/>
SELECT DISTINCT  ?mappingRule ?logicalSource ?predicate ?sourceAttribute
WHERE {
?mappingRule rml:logicalSource ?ls.
?ls          rml:source        ?logicalSource.
?mappingRule rr:subjectMap     ?subject.
?subject     rr:class          COVID-19:Publication.
OPTIONAL {   ?mappingRule rr:predicateObjectMap ?pObjectMap .
             ?pObjectMap  rr:predicate          ?predicate .
             ?pObjectMap  rr:objectMap          ?objectMap .
             ?objectMap   ?mode                 ?sourceAttribute}}     
    \end{lstlisting}
    \caption{SPARQL Query to retrieve the RML rules that define \texttt{COVID-19:Publication}}
    \label{fig:query1}
\end{figure}
\iffalse
\begin{listing}
  \caption{SPARQL Query to Retrieve the RML Mapping Assertions that define class Publication}
 \label{ex:query1}
 \begin{minted}{sparql}
PREFIX rr:  <http://www.w3.org/ns/r2rml#>
PREFIX rml: <http://semweb.mmlab.be/ns/rml#>
PREFIX COVID-19:  <http://research.tib.eu/covid-19/vocab/>
SELECT DISTINCT  ?mappingRule ?logicalSource ?predicate ?sourceAttribute
WHERE {
?mappingRule rml:logicalSource ?ls.
?ls          rml:source        ?logicalSource.
?mappingRule rr:subjectMap     ?subject.
?subject     rr:class          COVID-19:Publication.
OPTIONAL {   ?mappingRule rr:predicateObjectMap ?pObjectMap .
             ?pObjectMap  rr:predicate          ?predicate .
             ?pObjectMap  rr:objectMap          ?objectMap .
             ?objectMap   ?mode                 ?sourceAttribute}}  
\end{minted}
\end{listing}
\fi

\subsection{The Knowledge4COVID-19 KG in Numbers}
\begin{figure}[ht!]
\centering  
\hspace{0pt}{
     \includegraphics[width=1.0\textwidth]{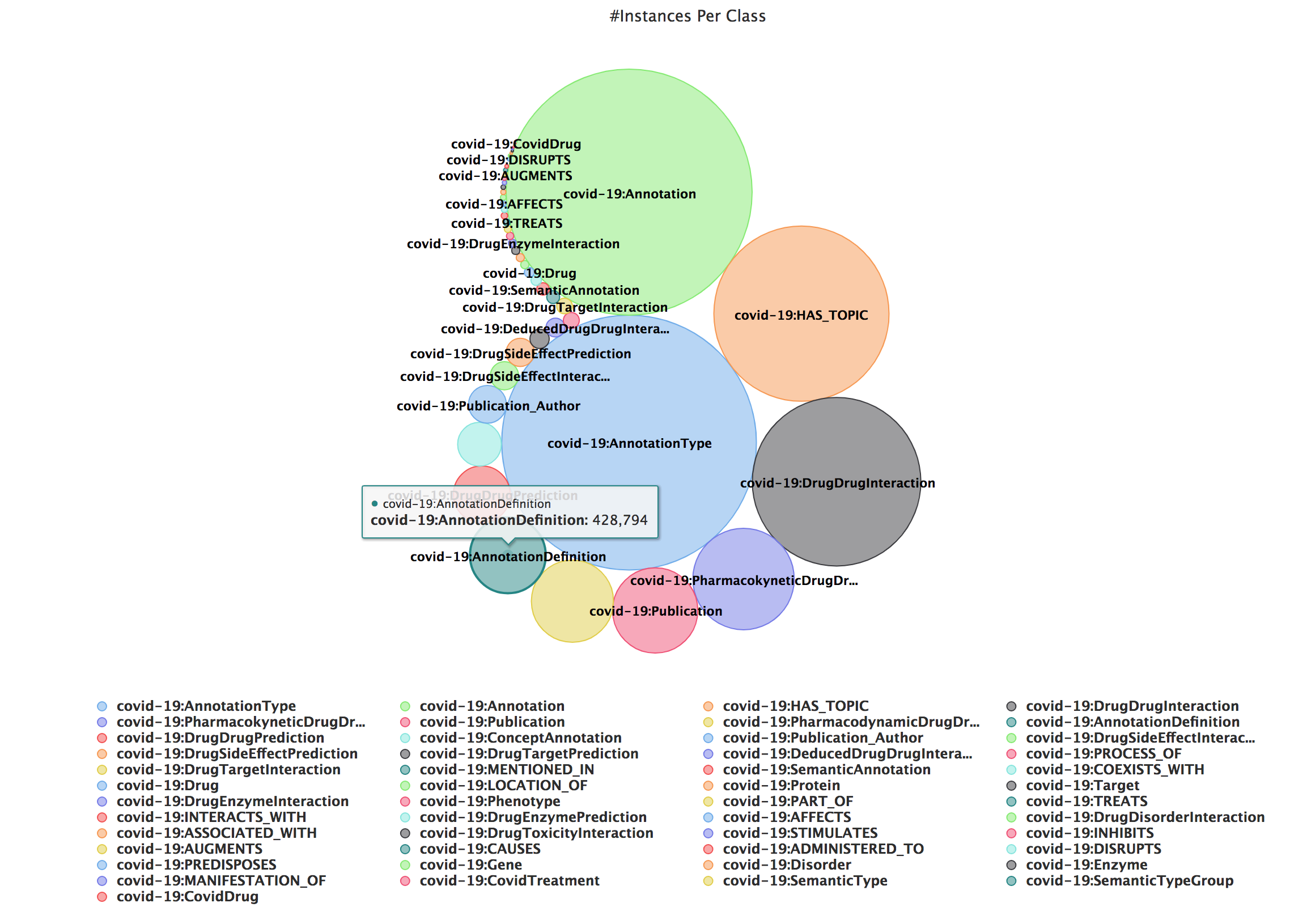}
    }
     \caption{\textbf{Knowledge4COVID-19 KG}. Number of Resources per classes; 4,864,162 annotations encode the meaning of 542,672 scientific publications and open data. } 
 \label{fig:rpc}
\end{figure}

  The current version of the Knowledge4COVID-19 KG comprises 80,570,440 RDF triples.
\autoref{fig:rpc} depicts the number of resources per class in the Knowledge4COVID-19 KG. As observed,  \texttt{covid-19:Annotation} comprises 4,536,579 resources, 542,672 resources in \texttt{covid-19:Publication}, 
503,700 for \texttt{covid-19:PharmacokyneticDrugDrugInteraction}. 
The Knowledge4COVID-19 KG includes 87 COVID-19 drugs; 68 drugs are from DrugBank\footnote{\url{https://go.drugbank.com/covid-19}} and the rest have been extracted from the Mayo Clinical website \footnote{\url{https://www.mayoclinic.org/diseases-conditions/coronavirus/expert-answers/coronavirus-drugs/faq-20485627}}. 
Additionally, the Knowledge4COVID-19 KG integrates 216 COVID-19 treatments that comprise COVID-19 drugs and drugs for the most common comorbidities that impact on the survival of COVID-19 patients \cite{covid19}: hypertension, depressive syndrome anxiety, obesity, cardiopathy, diabetes mellitus, hepatitis disease, chronic obstructive pulmonary disease, renal disease, asthma, dyslipidemia hypercholesterolemia, neurodegenerative disorder, gastrointestinal disease, vascular disease, benign prostatic hyperplasia, and obstructive sleep apnea. There are 923 deduced DDIs (a.k.a. DeducedDDIs).
In average, each COVID-19 treatment has 10.63 drugs, 1.58 COVID-19 drugs, and 9.11 comorbidity drugs. Additionally, COVID-19 treatments have in average two comorbidities and 121.43 DeducedDDIs; the same DDI can produce different effects, and they are counted as different DDIs. 
Moreover, the Knowledge4COVID-19 KG integrates 345,116 CRD and 5,513 NCRD pairs of drugs, and
124,537 instances of predicted DDIs  (i.e., instances of the class \texttt{covid-19:DrugDrugPrediction}). Specifically, 8,925 of the predicted DDIs are generated by the DDI-BLKG method, 5,907 have a score equal or greater than 0.5 (a.k.a. DDI-BLKG-0.5). The rest of the DDIs are discovered by state-of-the-art methods; they are included in the KG to provide a baseline for future benchmarking. These DDIs are predicted from the DDIs extracted from DrugBank, and are as follows:
\begin{inparaenum}[\bf i\upshape)]
%\item KBMF2K \cite{arxiv.1211.1275}: 263 DDIs predicted by the Kernelized Bayesian Matrix Factorization with twin Kernels. 
%\item LapRLS \cite{LapRLS}: 60,758 DDIs generated by Laplacian Regularized Least Squares.
%\item BLM \cite{BLM}: 81,418 DDIs generated by the Bipartite Local Method. 
%\item GIP \cite{btr500}: 6,584 DDIs generated by Gaussian Interaction Profile.
%\item DDI-LPB \cite{KritharaARNBVMG19}: 83,377 DDIs generated by machine learning methods that resort to knowledge encoded in scientific publications from PubMed to predict potential interactions between drugs.
\item TransE \cite{BordesUGWY13} 28,752 DDIs generated by TransE.  
\item RESCAL \cite{NickelTK11} 28,752 DDIs generated by RESCAL.  
\item HolE \cite{NickelRP16} 28,752 DDIs generated by HolE.  
\item DistMult \cite{arxiv.1412.6575} 28,752 DDIs generated by DistMult.  

\end{inparaenum}

\section{Exploring the Knowledge4COVID-19 KG}
\label{sec:exploration}
This section describes the services implemented to facilitate the traversal and data retrieval on top of the Knowledge4COVID-19 KG.

\subsection{Relevant Adverse Effects Detected on Knowledge4COVID-19}
\begin{figure*}[h!]
\centering
            \includegraphics[width=\textwidth]{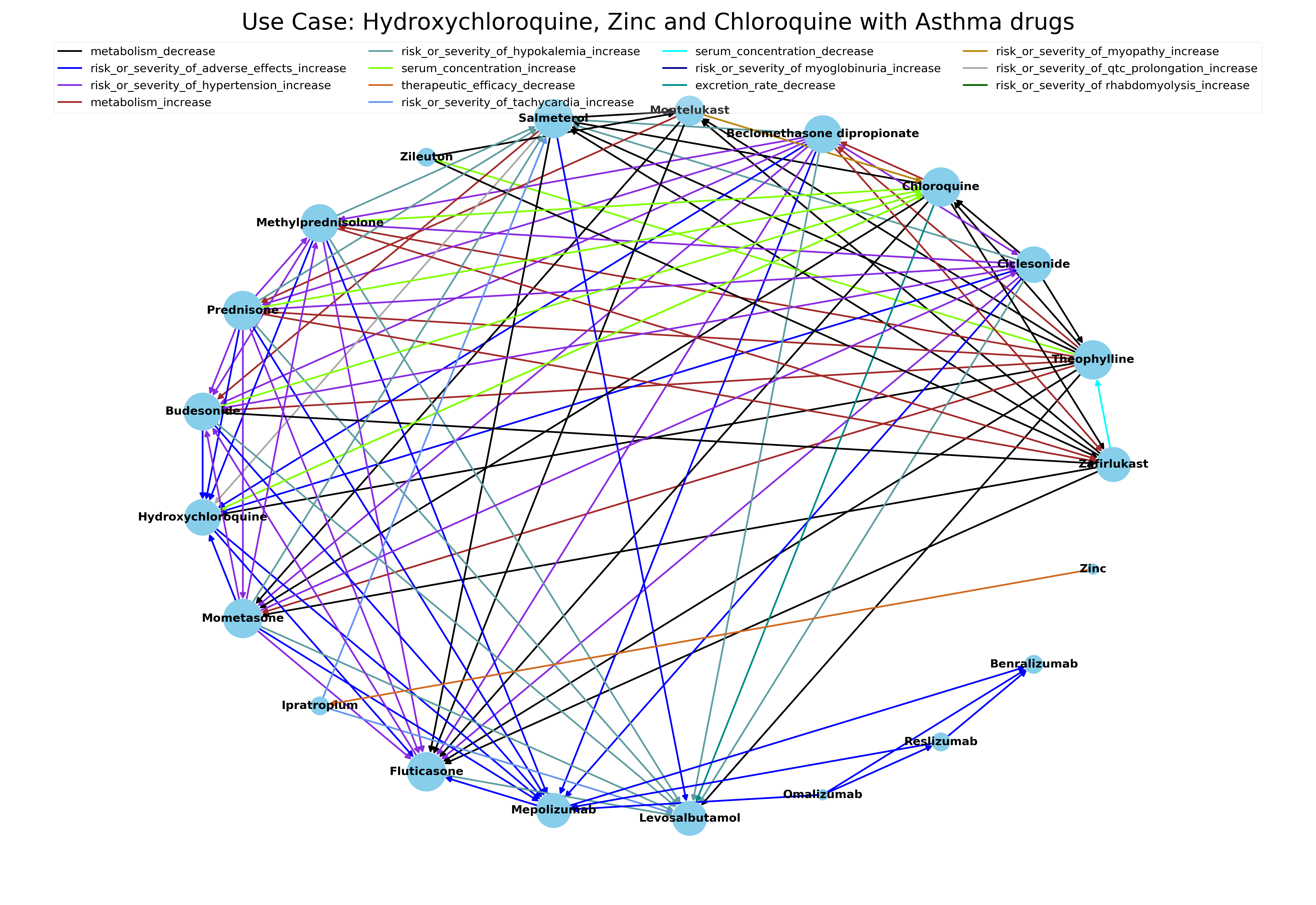}
            \caption{The adverse effects generated as the result of the interactions among COVID-19 drugs (Hydroxychloroquine, Zinc, and Chloroquine) with treatments for Asthma. Relations retrieved from the Knowledge4COVID-19 KG}
            \label{fig:asthma}
\end{figure*} 

This service aims at providing the support for analyzing relevant adverse effects that may be produced as a result of interactions among drugs to treat COVID-19 and conditions. As a proof of concept, we illustrate the results of the analysis of the most common commorbidities, i.e., hypertension, asthma, and diabetes. These comorbidities are linked to the ACE-2 receptor expression and may facilitate the entry of the virus into the host cells as a consequence of releasing the proprotein convertase. More importantly, this effect may fire a "vicious infectious circle" which may result in the increase of morbidity and mortality~\cite{covid19-2020}. Nevertheless, a more detailed analysis of the impact of the combination of drugs can be executed on the public available Jupyter Notebook\footnote{\url{https://colab.research.google.com/drive/146-oQTxDpZQoOifKY6iafaEwuupH7q3t?usp=sharing}}. Exemplar drug-drug interactions represented in the Knowledge4COVID-19 KG can also be visualized\footnote{\url{https://youtu.be/7YsTYJzRfR0}}.
 
\begin{figure*}[h!]
            \centering
            \includegraphics[width=\textwidth]{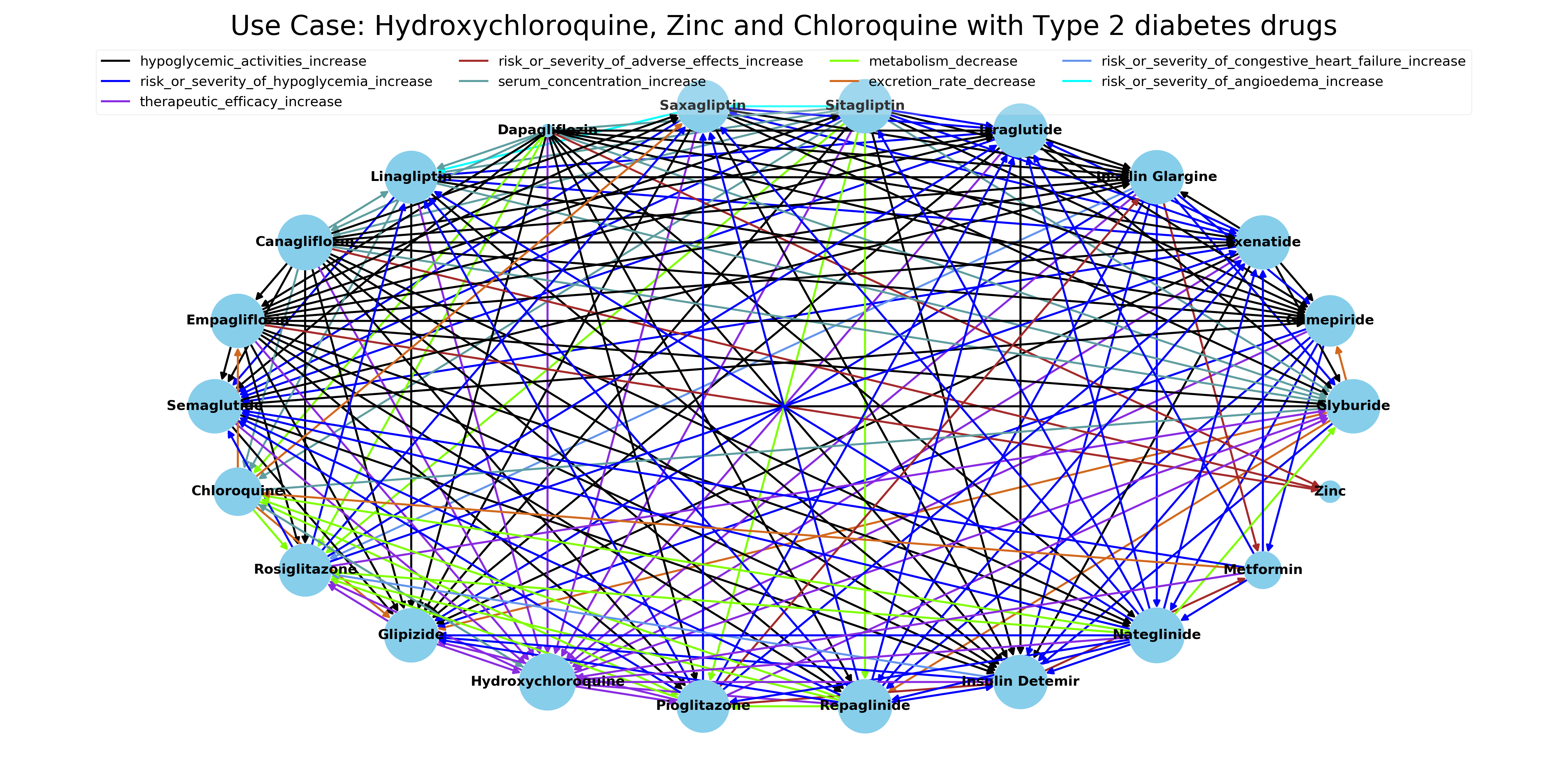}
            \caption{The adverse effects generated as the result of the interactions among COVID-19 drugs (Hydroxychloroquine, Zinc, and Chloroquine) with treatments for Type 2 Diabetes. Relations retrieved from the Knowledge4COVID-19 KG}
            \label{fig:diabetes}
\end{figure*} 

\begin{figure*}[h!]
            \centering
            \includegraphics[width=\textwidth]{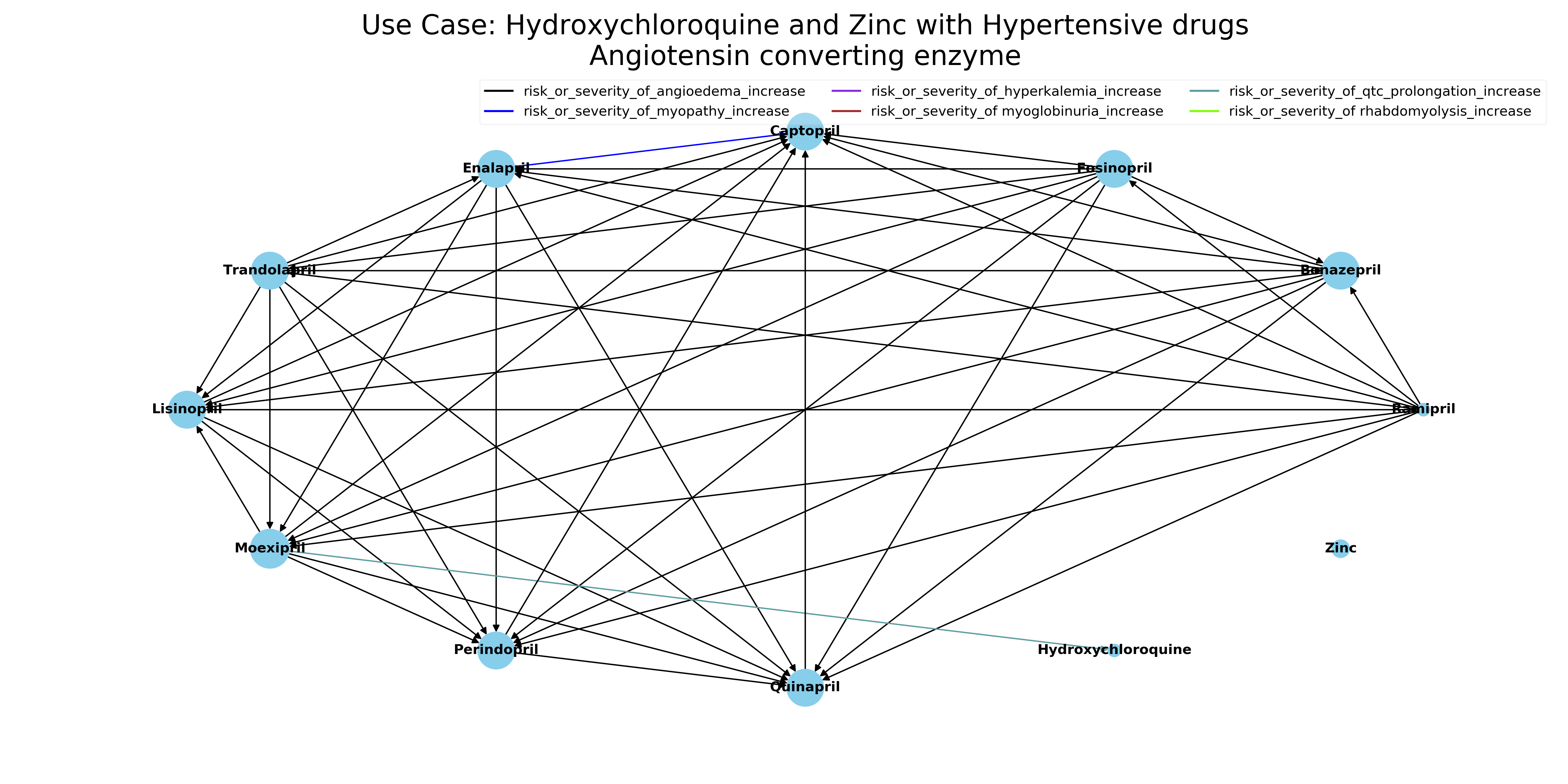}
            \caption{The adverse effects generated as the result of the interactions among COVID-19 drugs (Hydroxychloroquine, Zinc, and Chloroquine) with treatments for Hypertension. Relations retrieved from the Knowledge4COVID-19 KG}
            \label{fig:tension}
\end{figure*}

\autoref{fig:asthma}, \autoref{fig:diabetes}, and \autoref{fig:tension} depict adverse effects that can be triggered in COVID-19 patients who receive treatments for hypertension, asthma, or diabetes. Each plot reports a labelled directed graph, nodes represent drugs and an edge between two drugs, represent an interaction. The label of an edge, denoted by the line color and the figure legend, indicate the type of side effect. 

\autoref{fig:asthma} presents 14 types of drug-drug interactions that may occur among the COVID-19 drugs Hydroxychloriquine, Zinc, and Chloroquine, and asthma drugs. The pharmacokinetic drug-drug interactions between a pair of drugs A and B indicate that A impacts B's absorption, metabolism, excretion when both drugs are administrated together. As a result, A may reduce the effectiveness or increase toxicities.
The rest of the interactions are pharmacodynamic, i.e., their pharmacological outcome may be affected.
Six out of the 14 reported drug-drug interactions are pharmacokinetic. Chloroquine may reduce the metabolism of Zafirlukast, Mometasone, and Fluticasone; it can also decrease the excretion rate of Levosalbutamol. Hydroxychloriquine also impacts the metabolism of Theophylline. Furthermore, the serum concentration of Chloroquine may be increased with asthma drugs
by Methylprednisolone, Prednisone, and Budesonide. Thus, the effectiveness of the treatment was negatively affected.
Four drugs may increase the severity of the side effects of Hydroxychloriquine. At the pharmacodynamic level, it can be observed that Montelukast and Chloroquine may increase the risk of myopathy, and Salmeterol and Hydroxychloriquine may increase the risk of QT prolongation. Since the risk of cardiac events during QT syndrome is high, these results suggest that the combinations of the treatments need to be administrated with great precaution. 
Similarly, \autoref{fig:diabetes} reveals a more significant number of interactions among the drugs Hydroxychloriquine, Zinc, and Chloroquine and the drugs typically prescribed to Type 2 diabetes patients. All the drugs affect the efficacy of Hydroxychloriquine and the combination of Rosiglitazone in treatments with Insulin Determir or Insulin Glargine. Additionally, the therapeutic efficacy of Rosiglitazone can be increased when used in combination with Hydroxychloroquine, and Chloroquine may reduce the effectiveness of Metformin. They should be administrated with precaution because 
their therapeutic efficacy may be reduced. 
Drug interactions of hypertension treatments based on drugs Angiotensin converting enzyme, with the drugs Hydroxychloriquine and Zinc are reported in \autoref{fig:tension}.
As reported, the combination of these drugs may cause pharmacodynamic interactions that can critically affect the function of nerve and muscle cells, including those in the heart. 
The above results suggest that COVID-19 patients receiving treatments for pre-existing conditions need to be carefully treated. 

\subsection{Web APIs to traverse the Knowledge4COVID-19 KG}
The Knowledge4COVID-19 KG can be explored by executing SPARQL queries against the public SPARQL endpoint\footref{KGendpoint}. Additionally, specific Web Application Programming Interfaces (APIs)\footref{API} allow for the execution of specific requests. They include \begin{inparaenum}[\bf i\upshape)]
    \item the Publications related to drugs;
    \item the Drug-Drug Interactions between two or several drugs;
    \item the Predicted Drug-Drug Interactions between two or several drugs. 
    \end{inparaenum}
    The source code and the description of how to use the API is available on our GitHub repository\footref{API}. The Web APIs were executed 20 times, and the average execution time is reported. 
\\
\noindent
\textbf{Publications related to drugs} retrieves the scientific publications annotated with UMLS Concept Unique Identifiers (aka CUIs) of the input drugs.
\\
\noindent
\textit{Input:} CUI ids for one or several drugs.
\\
\noindent
\textit{Output:} All the properties of the publications annotated with input drugs.
\\
\noindent
\textit{Pre-conditions:} Publications are correctly annotated with CUIs.
\\
\noindent
\textit{Post-conditions:} Returned publications have mentions of the input drugs with respect to the CUI annotations in the abstract or title.
\\
\noindent
\textit{Average response time:} 50 ms.
\\
\noindent
\textit{Example SPARQL Query:} \autoref{appendix:publications}.
\\
\noindent
\textbf{Drug-Drug Interactions (DDI)} retrieves the DDI of the input drugs.
   
\noindent\textit{Input:} Drug CUIs and a variable ``\texttt{target}'' to indicate the output mode.

\noindent\textit{Output:} Drug-Drug Interactions related to the input drugs with all the properties defined in the KG. Interactions of the related drugs are returned as an output. Each interaction includes the effector drug, the affected drug, the effect, and the impact of the effect. If the variable \texttt{target}=DDI, then return the DDI of each input drug individually. If \texttt{target}=drug-drug interactions, then return the DDI of all the possible pairs of the input drugs.

\noindent\textit{Pre-conditions:} Drugs have interactions in the KG; these interactions are extracted from DrugBank or the literature. 

\noindent\textit{Post-conditions:} Returned interactions are related to the drugs in the input.
\\
\noindent
\textit{Average response time:} 62 ms.
\\
\noindent
\textit{Example SPARQL Query:} \autoref{appendix:interactions_drug} and \autoref{appendix:interactions_drugs}.
\\
\noindent
\textbf{Predicted Drug-Drug Interactions} retrieves predicted DDI of input drugs.
   
\noindent\textit{Input:} Drug CUI and a variable ``\texttt{target}'' to indicate the output mode.

\noindent\textit{Output:} Predicted Drug-Drug Interactions related to the input drugs with all the properties defined in the KG. Predicted interactions of the related drugs are returned as an output. Each interaction consists of the effector drug, the affected drug, a confidence score of the interaction, and the provenance. If \texttt{target}=DDIP, then the API returns the predicted DDI of each drug individually. If \texttt{target}=DDIPS, then the API returns the predicted DDI of all the possible pairs of the input drugs.

\noindent\textit{Pre-conditions:} Drugs have predicted interactions in the KG.

\noindent\textit{Post-conditions:} Returned predicted interactions have a confidence score greater than zero wrt the CUI of the drugs in the input.

\noindent\textit{Average response time:} 58 ms.

\noindent
\textit{Example SPARQL Query:} \autoref{appendix:predicted_interactions_drug} and \autoref{appendix:predicted_interactions_drugs}.

\subsection{Federated Query Processing on top of the Knowledge4COVID-19 KG}
\begin{figure}[t]
    \begin{lstlisting}[style=sparql,basicstyle=\footnotesize]
PREFIX dbp: <http://dbpedia.org/property/>
PREFIX owl: <http://www.w3.org/2002/07/owl#>
PREFIX k4covid: <http://research.tib.eu/covid-19/vocab/>
PREFIX k4covide: <http://research.tib.eu/covid-19/entity/>

SELECT DISTINCT * WHERE {
  SERVICE <https://labs.tib.eu/sdm/covid19kg/sparql> {
    ?treatment k4covid:hasCovidDrug ?covidDrug.
    FILTER( ?comorbidity=k4covide:Asthma )
    ?treatment k4covid:hasComorbidity ?comorbidity.
    ?treatment k4covid:hasComorbidityDrug ?comorbidityDrug.
    ?comorbidityDrug k4covid:hasCUIAnnotation ?CUIComorbidityDrug.
    ?CUIComorbidityDrug owl:sameAs ?sameAsComorbidityDrug .
    ?covidDrug k4covid:hasCUIAnnotation ?CUICovidDrug.
    ?CUICovidDrug owl:sameAs ?sameAsCovidDrug .
  }
  SERVICE <https://dbpedia.org/sparql> {
    ?sameAsCovidDrug dbp:excretion ?excretation.
    ?sameAsCovidDrug dbp:metabolism ?metabolism.
    ?sameAsCovidDrug dbp:routesOfAdministration ?routes.
  }  
}
    \end{lstlisting}
    \caption{Example of a Federated Query}
    \label{fig:federated-query}
\end{figure}

DeTrusty~\cite{DeTrusty} is a federated query engine for RDF sources.
Hence, it allows querying the Knowledge4COVID-19 KG in conjunction with external sources like DBpedia, Wikidata, and Uniprot\footnote{\url{https://labs.tib.eu/sdm/k4covid-query-engine/sparql}}.
This in turn is only possible because entities in the Knowledge4COVID-19 KG are linked to those datasets.
\autoref{fig:federated-query} shows an example of a federated query; providing information about treatments that involve drugs for COVID-19 and Asthma.
DeTrusty contacts both KGs to retrieve the complete answer to the query.
The Knowledge4COVID-19 KG delivers data about the treatments fulfilling the conditions; including \texttt{owl:sameAs} links for both drugs.
DeTrusty also contacts DBpedia to get additional information about the COVID-19 drugs, e.g., the route of administration.
DeTrusty decomposes the SPARQL query into star-shaped sub-queries around the subjects~\cite{VidalRLMSP10}, i.e., each triple of a sub-query has the same variable or constant in the subject position.
For source selection in the presence of a SPARQL query without the \texttt{SERVICE} clause, DeTrusty uses a semantic source description with information about the classes and their predicates, like MULDER~\cite{Endris2018}.

\section{Evaluation of Knowledge4COVID-19}
\label{sec:eval}
In this section, we report on the evaluation of the quality of the integrated data and the patterns discovered by exploiting the knowledge encoded in the Knowledge4COVID-19 KG. We aim to answer the following research questions:
\begin{inparaenum}[\bf {\bf Q}1\upshape)]
\item What is the accuracy of the named entity recognition (NER) and named entity linking (NEL) performed over data from DrugBank to extract drug-drug interactions and the effects of these interactions?
\item What is the accuracy of the prediction methods that enhance the knowledge about drug-drug interactions?
\item What is the quality of the knowledge discovery methods implemented on top the Knowledge4COVID-19 KG to uncover drug-drug interactions among the multi-drug COVID-19 treatments?
\end{inparaenum}

\subsection{Effectiveness of NER and NEL methods}
Data about drug-drug interactions is collected from DrugBank release 2022-01-04 with 1,273,052 entries composed of pairs of drugs and the textual description of the effects of each interaction. 
In order to evaluate the performance of FALCON in this use case, 1,198 DDI descriptions were manually annotated by twelve annotators; annotations correspond to CUIs from UMLS and constitute the gold standard of the evaluation. For example, for the DDI description: “The serum concentration of Lepirudin can be decreased when it is combined with Tipranavir”; Lepirudin and Tipranavir correspond to the extracted entities from the above record, while decrease and serum concentration represent, respectively, the effect and impact of the interaction of Tipranavir with Lepirudin. One of the annotators was a senior researcher, two were experts in the biomedical domain, and the rest were Computer Science PhD students. Disagreements among the annotators were solved by majority voting. A 2-fold cross-validation was followed. The evaluation indicates a precision of 98\%. The 2\% where FALCON failed to extract and link the terms correctly, are interactions which contain more than one interaction in the same sentence, where FALCON was only considering one interaction (\autoref{table:ddi_patterns} last pattern). All the drug-drug interactions, that followed this pattern, were corrected manually before integrating them into the Knowledge4COVID-19 KG.

\subsection{Effectiveness of the predictive tasks for DDI identification}
\label{sec:pred}
As explained, DrugBank presents an adequate source for retrieving potential adverse effects of treatments received by COVID-19 patients in combination with other medications, explicitly providing the interactions of each drug with other drugs in a structured way. However, many drug interactions observed in everyday practice are not currently recorded in medical databases like DrugBank that are continuously evolving and extended with new drugs and relevant information\footnote{\url{https://go.DrugBank.com/release\_notes}}. Thus, a Knowledge Graph completion challenge arises, for adding new drug-drug interaction links in the Knowledge4COVID-19 KG.

\begin{figure*}[ht!]
\centering
{\includegraphics[width=1.0\linewidth]{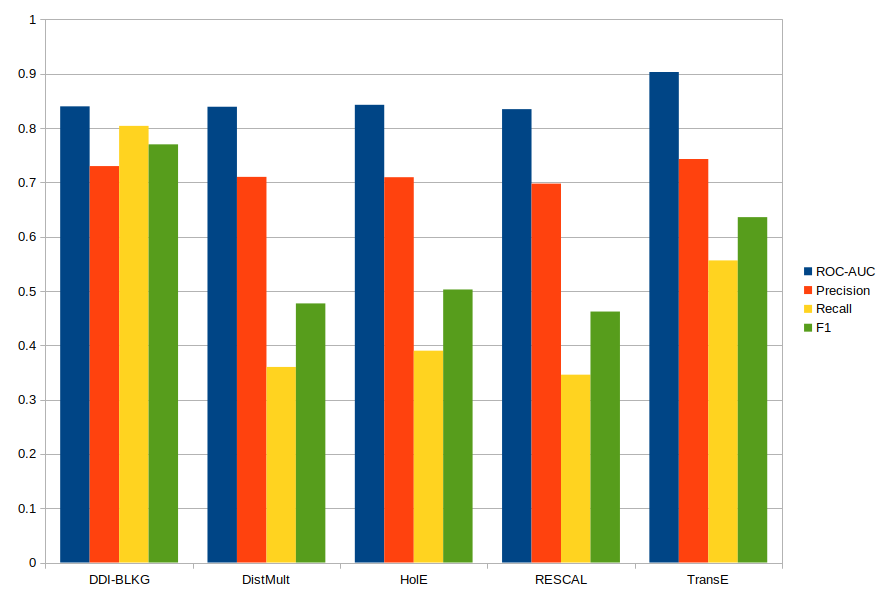}} 
\caption{ Results of the 10-fold cross validation, comparing the current DDI prediction approach (DDI-BLKG) with various graph embedding methods. Only the TransE approach outperforms our approach in ROC-AUC and Precision, while in terms of the F1 score, DDI-BLKG outperforms all embeddings by far.}
\label{fig:evaluation}
\end{figure*}

The effectiveness of the Random Forest classifier presented previously in \autoref{sci-pub-de}, for predicting new interactions is assessed following a 10-fold cross-validation (cv) procedure, and DrugBank v5.0.3 is our gold dataset for drug-drug interactions. Existing techniques for knowledge graph embeddings available in the TorchKGE \footnote{\url{https://torchkge.readthedocs.io}} library (i.e., TransE, RESCAL, HolE, and DistMult) are used as baselines. Each model is trained for a maximum of 100 epochs while early stopping was used,
utilizing 10\% of the data for validation. For each model, 100-sized embeddings were used, since an increase of the embedding size did not provide better results. The performance of the predictive models is measured using the area under the receiver-operating characteristic (ROC-AUC), as well as the macro-average of Precision, Recall and F1-score for the positive class. \autoref{fig:evaluation} suggests that our approach (DDI-BLKG) outperforms all mainstream embedding-based methods tested. DDI-BLKG can exploit knowledge encoded in the fine-grained representation of the publications in the Knowledge Graph. As a result, the DDI-BLKG prediction accuracy is enhanced compared to the baseline methods.
\\ 

\begin{figure*}[t!]
\centering
{\includegraphics[width=1.0\linewidth]{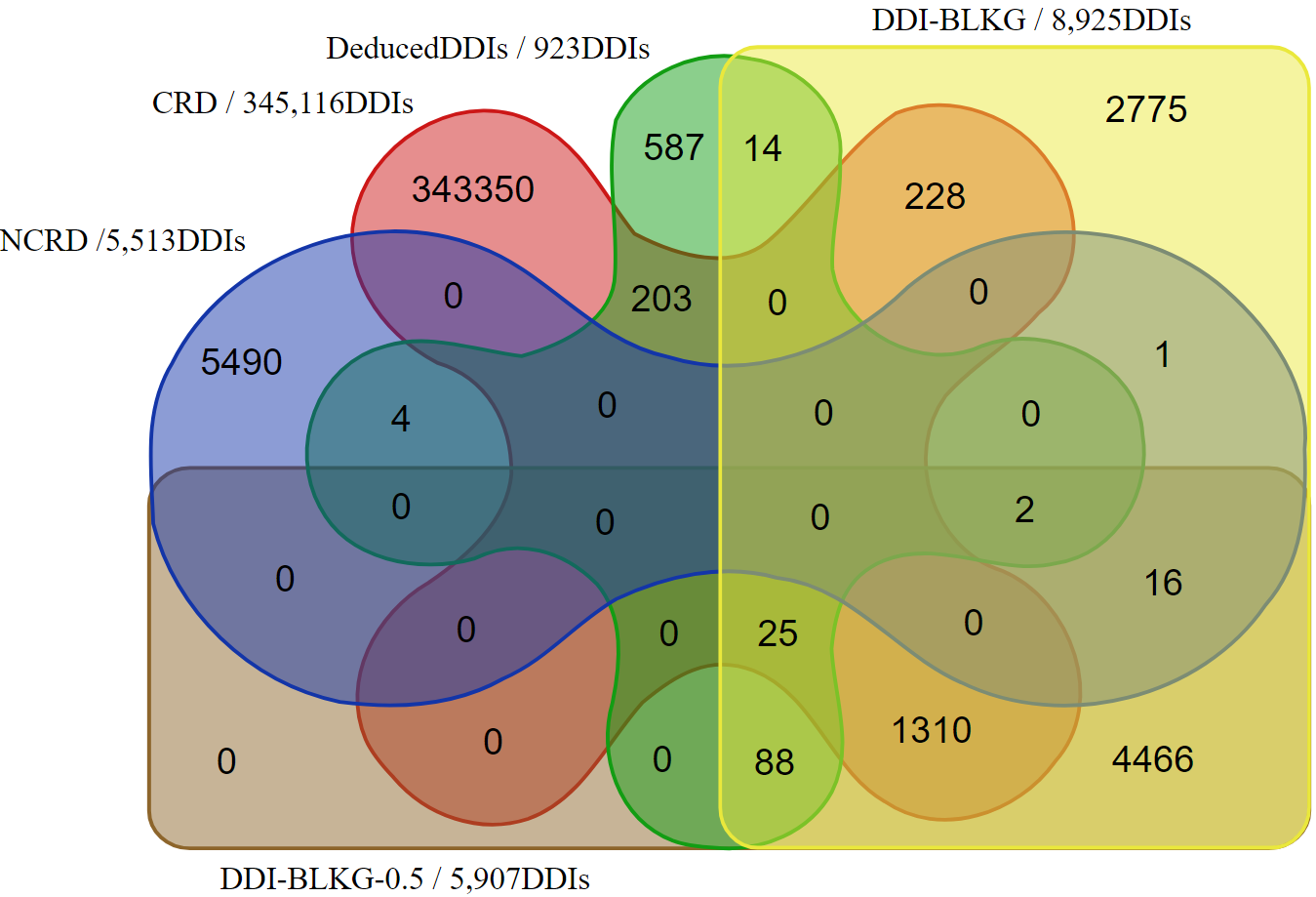}} 
\caption{Venn Diagram. It depicts the overlap among five sets of DDIs. 345,116 CRD pairs of drugs targeting at least one protein of the family CYP. 5,513 NCRD are pairs of drugs targeting a No CYP protein. 8,925 DDI-BLKG are DDIs predicted by the DDI-BLKG method, while 5,907 DDI-BLKG-05 represents the subset of DDIs in DDI-BLKG with score equal or greater than 0.5. 923 DeducedDDIs generated by the deductive system. }
\label{fig:venn_diagram}
\end{figure*}

Moreover, \autoref{fig:venn_diagram} reports on the overlap between the DDIs deduced on the drugs of the COVID-19 treatments (a.k.a. DeducedDDIs), DDI-BLKG, DDI-BLKG-0.5 (DDI-BLKG with a prediction score equal or greater than 0.5), CRD, and NCRD. 
It is essential to highlight that CRD and NCRD are computed from the whole DrugBank dataset of drugs, while DDI-BLKG and 
DeducedDDIs are limited to COVID-19 drugs. The percentages of overlap of DeducedDDIs, DDI-BLKG, and DDI-BLKG-0.5 with CRD are 24.70\%, 17.51\%, and 22.60 \%. Thus, both methods (i.e., the deductive system and DDI-BLKG) can identify DDIs between drugs mediated by the CYP enzyme family, i.e., CRD pairs of drugs. CYP enzymes play an important role in catalyzing the metabolism of pharmaceuticals and their inhibition or induction causes clinically significant pharmacokinetic drug-drug interactions \cite{Hakkola}. Thus, these results suggest that even though these methods do not exploit any information about the drug's target enzymes, they can identify a good proportion of DDIs that are part of the CRD group. 

\subsection{Impact on the Effectiveness and Toxicity of COVID-19 Treatments}
\label{sec:toxicity}
The Knowledge4COVID-19 KG is a unique source of knowledge to identify patterns in the integrated networks of interactions, biomedical entities, and publications, e.g. adverse events generated by combining COVID-19 drugs and drugs prescribed for pre-existing conditions. Note that existing tools (e.g., COVID-19 Drug Interactions for University of Liverpool\footnote{\url{https://www.covid19-druginteractions.org/}}) only identify pair-wise interactions. In this section, we evaluate the drug-drug interactions that can be deduced over the Knowledge4COVID-19 KG and the effects of these interactions.
We consider five COVID-19 treatments and the effects of including in these treatments drugs for comorbidities. The treatment for COVID-19 used in these five cases is recommended by the official guidelines\footnote{\url{https://www.covid19treatmentguidelines.nih.gov/therapies/antiviral-therapy/chloroquine-or-hydroxychloroquine-and-or-azithromycin/}}.
The concomitant drugs used in the first treatment $T1$ are for the comorbidities asthma, high cholesterol, and pneumonia and for the second treatment $T2$ are diabetes, hypertension, and pneumonia. The comorbidities in the third treatment $T3$ are diabetes, high cholesterol, hypertension. The comorbidities in the fourth treatment $T4$ are asthma and hypertension, and for the fifth treatment, $T5$ are renal diseases, obesity, and hypertension.

\autoref{experimental_study} shows the percentage of DDIs deduced ($D$) and wedge absolute frequency ($F$) for each middle-vertex by the method~\cite{DBLP:conf/kcap/RivasV21} in existing treatments.
\begin{table*}
\begin{center}
\caption{\textbf{Five COVID-19 Treatments}. Frequency distribution of wedges with knowledge capture. Treatments are evaluated in four interaction checker tools: COVID-19, WebMD, Medscape, and DrugBank (May 2nd, 2022). For each tool, it is shown the DDI-Reduction percentage that indicates how many DDIs are avoided in a treatment when the middle-vertex drug is removed. The DDI-reduction percentage is a higher-is-better metric. Middle-vertex drugs reduce DDIs, suggesting, thus, wedges and their middle vertices are part of DDIs that affect treatment effectiveness and toxicities. Best values in \textbf{bold}.} \label{experimental_study}
\scriptsize{
\begin{tabular}{c|cc|c|cccc}
\toprule 
\textbf{T} &  \multicolumn{2}{c|}{\textbf{Knowledge Capture}} & $D$ & \multicolumn{4}{c}{\textbf{DDI-Reduction Percentage}}\\\cmidrule{2-3}\cmidrule{5-8}

& Middle-Vertex & $F$ &  & COVID-19 & WebMD & Medscape & Drugbank \\
\toprule
T1 &  \textbf{Azithromycin} & \textbf{9} & 45.45 & \textbf{100.0} & \textbf{100.0} & \textbf{100.0} & 42.9\\
    &   Montelukast    		& 4 &  \\
    &   Lovastatin     		& 4 &  \\
    &   Hydroxychloroquine	& 0 &  \\
    &   Doxycycline			& 0 &  \\\cmidrule{1-8}
 
T2 &  \textbf{Ciprofloxacin} & \textbf{12} & 52.17 & 33.3 & \textbf{75.0} & \textbf{75.0} & 44.4\\
    &   \textbf{Metoprolol}    & \textbf{12} &  & 33.3 & 25.0 & 25.0 & 33.3\\
    &   Hydroxychloroquine     & 9 &  \\
    &   Azithromycin      & 9 &  \\
    &   Linagliptin     & 7 &  \\\cmidrule{1-8}
	
T3 &  \textbf{Hydroxychloroquine} & \textbf{5} & 33.33 & \textbf{100.0} & 25.0 & 25.0 & 60.0\\
    &   \textbf{Glyburide}    & \textbf{5} & & 0.0 & 50.0 & 50.0 & 60.0\\
    &   Simvastatin      & 3 &  \\
    &   Azithromycin       & 3 &  \\
    &   Ramipril     & 0 &  \\\cmidrule{1-8}

T4 &  \textbf{Propranolol} & \textbf{8} & 15.38 & \textbf{100.0} & 50.0 & 50.0 & 60.0\\
    &   Hydroxychloroquine    & 5 &  \\
    &   Azithromycin      & 5 &  \\
    &   Theophylline       & 4 &  \\
    &   Ramipril     & 1 &  \\\cmidrule{1-8}

T5 &  \textbf{Timolol} & \textbf{11} & 38.89 & \textbf{50.0} & \textbf{50.0} & \textbf{50.0} & 44.4\\
    &   \textbf{Cyclophosphamide}    & \textbf{11} &  & 0.0 & 0.0 & 0.0 & 44.4\\
    &   Azithromycin      & 7 &  \\
    &   Hydroxychloroquine       & 7 &  \\
    &   Bupropion     & 6 &  \\
\bottomrule
\end{tabular}
}
\end{center}
\end{table*}

The middle-vertex of a wedge is highly important because the middle-vertex is both the object drug for one interaction and the precipitant drug for another interaction.
Thus, drugs that correspond to the middle-vertex of wedges, represent drugs whose presence in the treatment may negatively impact effectiveness and toxicity. We can observe in \autoref{experimental_study} that over 15\% of new DDIs are deduced in all the treatments.
\autoref{experimental_study} shows the DDI-Reduction percentage for the drugs with higher wedge absolute frequency (F) for each treatment. The DDI-Reduction percentage was evaluated in four interaction checker tools on May 2nd, 2022, Liverpool COVID-19 Interactions\footnote{\url{https://www.covid19-druginteractions.org/checker}}, WebMD\footnote{\url{https://www.webmd.com/interaction-checker/default.htm}}, Medscape\footnote{\url{https://reference.medscape.com/drug-interactionchecker}}, and Drugbank\footnote{\url{https://go.drugbank.com/drug-interaction-checker}}. The validation was done on the versions of  Liverpool COVID-19 Interactions and Drugbank which correspond to 2022-04-13 and 2022-01-04, respectively.
DDI-Reduction percentage is measured, and it indicates how many DDIs are avoided in a treatment when the middle-vertex drug is withdrawn.
The evaluation suggests that withdrawing the middle-vertex with higher absolute frequency reduces most interactions.
Thus, wedges and their middle-vertex represent DDIs that affect treatment effectiveness and toxicities.
When more than one drug contains the higher wedge absolute frequency (F) in treatment, the clinicians have to decide which drug is withdrawn.
The first COVID-19 treatment reported contains concomitant drugs for the comorbidities of asthma, high cholesterol, and pneumonia. The method proposed by~\cite{DBLP:conf/kcap/RivasV21} indicates Azithromycin as the drug with the highest absolute frequency of being the wedges middle-vertex.
Therefore, it represents the DDIs that affect treatment effectiveness and toxicities, and withdrawing it reduces most interactions.

\section{Related Work}
\label{sec:rw}
\noindent\textbf{Data Ecosystems and Spaces}
The International Data Space (IDS)~\cite{DBLP:conf/semweb/BaderPMTQMABILG20} exemplifies DEs where various W3C standards, technologies, and governance models allow for the description of the data sources to secure and standardize data exchange and integration.
Data ecosystems provide reference architectures that comprise components to enable the description of the data sources to be exchanged and mappings between data sources with integrated views or unified schemas. Specifically, the networks of knowledge-driven data ecosystems (by Geisler and Vidal, et al.~\cite{abs-2105-09312}) enable the nested definition of data ecosystems in terms of other data ecosystems whose connections induce a network. Metadata of each data ecosystem is described using controlled vocabularies and domain ontologies. Additionally, services are part of data ecosystems and can exploit metadata to enhance interoperability, traceability, data quality assessment, and integrity constraint validation. The Knowledge4COVID-19 data ecosystem implements the reference architecture proposed by Geisler and Vidal, et al.~\cite{abs-2105-09312}; it comprises the data ecosystem for Scientific Open Data and Scientific Publications. Mappings between data sources and the unified schema are defined using RML, and the execution of these mappings results in the materialized Knowledge4COVID-19 KG. Services for knowledge extraction and prediction are implemented at each data ecosystem. A deductive system, developed on top of the Knowledge4COVID-19 KG, facilitates discovering new drug-drug interactions and their effects on treatment toxicities and effectiveness. 

\noindent\textbf{Knowledge Graphs} have gained momentum as data structures able to model the convergence of data and knowledge as factual statements~\cite{DBLP:journals/cacm/GutierrezS21}. Despite being coined by the research community for several decades, KGs are playing an increasingly relevant role in scientific and industrial areas~\cite{NoyGJNPT19}. The research community has actively contributed to the problem of automatic knowledge graph creation. As a result, declarative specification of a KG creation process~\cite{iglesias2020sdm,JozashooriCIVC20,DBLP:journals/semweb/Chaves-FragaRPV21}, techniques for semantic data integration~\cite{Cudre-Mauroux20,DBLP:journals/dbsk/VidalEJSR19}, and virtualization of KGs~\cite{XiaoDCC19,DBLP:conf/dexa/EndrisRVA19} enable to merge data silos and provide an integrated view of data and metadata. 
Existing KG construction methods vary from crowdsourced (e.g., Wikidata~\cite{DBLP:journals/cacm/VrandecicK14}), extraction from existing knowledge bases (e.g., DBpedia~\cite{DBLP:conf/semweb/AuerBKLCI07} and YAGO~\cite{DBLP:journals/ai/HoffartSBW13}), and automatic generation (e.g., KnowledgeVault~\cite{knowledgevault} and AI-KG~\cite{DessiORBMS20}). Moreover, KG refinement includes methods for predicting relations, completing type assertions, and finding erroneous relations, external links, and values~\cite{Paulheim17}. 
The creation process of Knowledge4COVID-19 KG is declarative using RML mapping rules, facilitating, thus, extensibility, modularity, and reusability of the KG creation process. 

\noindent\textbf{Knowledge Graphs and COVID-19:}
Several authors have proposed using knowledge graphs to make available expressive sources of data and knowledge about COVID-19. 
Specifically, several knowledge bases have been developed to provide an integrated view of COVID-19-related data. Exemplar approaches include COVID-19 Knowledge Graph\footnote{\url{https://covidgraph.org/}},  Drugs4Covid~\cite{abs-2012-01953}. Similarly, 
Knowledge4COVID-19 integrates CORD-19 scientific publications, but in addition, it models a fine-grained representation of drug-drug interactions and their adverse effects in the treatments of comorbidities. 
Additionally, Queralt-Rosinach et al.~\cite{Queralt-Rosinach} present a knowledge graph that integrates clinical data collected in the context of the BEAT-COVID project \footnote{\url{https://www.izi.fraunhofer.de/en/about-us/united-against-corona/beat-covid.html}}. These approaches put in perspective the protagonist role of knowledge graphs in understanding COVID-19.  
Similarly, Knowledge4COVID-19  aims to provide a resource that clinicians and patients can explore to understand the effects of interactions in a COVID-19. Thus, given the impact that pre-existing conditions seem to have on the outcome of a SARS-CoV-2 infection, Knowledge4COVID-19 represents a resource that can be linked to existing COVID-19 knowledge graphs to empower their analytical capacity.

Chatterjee et al.~\cite{chatterjee2021knowledge} present an exploratory review of recent works constructing knowledge graphs from different sources. For instance, Wang et al.~\cite{wang2020covid} have produced a literature knowledge graph construction and drug repurposing approach, also working on the fine-grained text entity extraction, while more recently authors in~\cite{domingo2021covid} also construct a knowledge graph from scientific literature, focusing on cause-and-effect relations. Knowledge4COVID-19 follows the best practices of FAIR~\cite{wilkinson2016} and Linked Data principles\footnote{\url{https://www.w3.org/TR/ld-bp/}}, and makes available a KG that integrates COVID-19 related data from various sources. UMLS is used to annotate biomedical entities; links to KGs (e.g., DBpedia and Wikidata) enhance interoperability. 

Reese et al.~\cite{DBLP:journals/patterns/ReeseUCCRCSGBFB21} describe a knowledge graph for COVID-19 where biomedical concepts and publications are represented at symbolic and subsymbolic levels. Complementary, Knowledge4COVID-19 provides a fine-grained representation of biomedical concepts and publications. Well-known tools like MetaMap and SemRep are used to extract relevant biomedical entities and relations from scientific publications. At the same time, drug indications, side effects, and adverse events of drug-drug interactions are recognized by FALCON2.0. The extracted entities are linked to equivalent resources in existing KGs (i.e., DBpedia, Wikidata, DrugBank, and Uniprot) and annotated using UMLS terms and relations; networks of drug-drug, drug-target, and drug-side effect interactions predicted using diverse methods are also merged. This makes the Knowledge4COVID-19 KG a complementary source of knowledge that can be connected to existing COVID-19 KGs (e.g., the one implemented by Reese et al.) using the linking techniques implemented by FALCON2.0. 

%\section{Discussion}.

\section{Knowledge4COVID-19 as a Resource}
\label{sec:resource}
\subsection{Discussion of the Knowledge4COVID-19 Framework}
This section describes our resources and discusses our contributions: 
\\
 \noindent\textbf{The Knowledge4COVID-19 DE} integrates data sources from the Scientific Open Data and Publications DEs. The pipeline for KG creation and management is available as a Docker container. It includes the Knowledge4COVID-19 unified schema, the RML triple maps, and the data sources processed by the NLP tools implemented by Scientific Open Data and Publications DEs. In addition, to create the Knowledge4COVID-19 KG, the pipeline uploads the KG to a Virtuoso endpoint and makes each resource available in the Knowledge4COVID-19 KG, following the Link Data principles. Moreover, the required configurations of the federated query engine DeTrusty are generated. DeTrusty is also available via its HTTP API like a regular SPARQL endpoint.   
\\
\noindent\textbf{The Knowledge4COVID-19 KG} comprises COVID-19 related data about drugs, DDIs (predicted and known), scientific publications, drugs' side effects, and interactions with targets. The KG can be explored through three APIs, a SPARQL endpoint, and a federated query engine. 
\\
\noindent\textbf{DDI Prediction Methods} employ machine learning techniques to identify previously unknown potential COVID-19 related drug-drug interactions with a certain confidence score. Predicted DDIs are not documented in open drug databases, such as Drugbank, and clinicians can use them as an indication of possible toxicities, during the treatment of a patient suffering from COVID-19.
\\
\noindent\textbf{Benchmarks of DDIs} include known, deduced, and predicted DDIs. The known DDIs are extracted from DrugBank, while CRD, NCRD, and DeducedDDIs are deduced. Finally, a set of DDIs predicted by state-of-the-art machine learning methods is also part of the Knowledge4COVID-19 KG. These DDIs can be used to reproduce our reported results or for future comparisons.

 \subsection{The Knowledge4COVID-19 Resource Characteristics}
\noindent\textbf{Novelty:} 
Knowledge4COVID-19 introduces a novel infrastructure to transform heterogeneous data sources into a KG. The mappings among the data sources and the unified schema are defined as RML mapping assertions. Moreover, the methods implemented in SDM-RDFizer allow for the efficient execution of the KG creation process. The Knowledge4COVID-19 KG occupies 13 GB and is created from 2.8 GB of raw data. Knowledge4COVID-19 KG executes 57 RML triples maps (comprising 223 mapping assertions) over the raw data in 82 minutes 55 seconds. Additionally, novel prediction methods are utilized to predict interactions between drugs.
We hope that these results encourage the community to create declarative pipelines for KG creation that are able to scale up to the avalanche of data expected in the next years. 
\\
\noindent\textbf{Availability:} Knowledge4COVID-19 is released publicly by the Scientific Data Management (SDM) group at TIB, Hannover. 
TIB is one of the largest libraries for science and technology in the world. Following its policy of engaging open access to scientific artifacts, it will support Knowledge4COVID-19 as a source of knowledge for SARS-CoV-2 and other viruses.
The \mbox{Knowledge4COVID-19 DE} is open source, written in Python 3, and uses RML; it is available under the Apache License 2.0 license in an open GitHub repository\footref{github}. It will be regularly updated with new data sources, triples maps, and APIs for exploration. More importantly, respecting open science good practices, Knowledge4COVID-19 is registered at Zenodo. Thus, users can use and cite a specific version, ensuring reproducibility and traceability of any experimental evaluation.
\\
\noindent\textbf{Utility:}
A docker image of Knowledge4COVID-19 is available at\footnote{\url{https://github.com/SDM-TIB/Knowledge4COVID-19/wiki/Running-Knowledge4COVID-19-KG-locally}}; it enables accessing the KG locally. The GitHub repository of the resource provides a detailed explanation of how to run the Docker container. From 24 to 26 April 2020, Knowledge4COVID-19 participated in the Pan-European hackathon \#EUvsVirus\footnote{\url{https://www.euvsvirus.org/}}
 organized with the aim of connecting experts, investors, and civilian organizations to devise together innovative solutions to the coronavirus outbreak\footnote{\url{https://devpost.com/software/COVID-19-kg}}. 
\\
\noindent\textbf{Predicted Impact:}
Open pharmaceutical databases such as Drugbank or drugs.com are periodically updated, manually adding drug-drug interactions, since new unknown DDIs are frequently reported by clinicians and health institutions. Our methods can potentially deduce or predict such interactions for new or experimental drugs by analyzing contextual information in biomedical publications before being observed in practice and documented. This could support treatment decision-making, avoiding unnecessary side effects of drug combinations.
Moreover, given the number of scientific publications and open data about drugs, disorders, and adverse events integrated into the Knowledge4COVID-19 KG, we are optimistic that it will be the starting point of future developments and benchmarking in the Semantic Web community. Lastly, the pipeline for KG management is domain agnostic, but there are still many opportunities to make it fully transparent. We hope this paper encourages the community to develop traceable and interpretable methods for transparent KG management. 
 \\
\noindent\textbf{Adoption and Reusability:}
We are reusing the same DE and core concepts of the unified schema and mapping rules in projects like EU H2020 projects like iASiS\footnote{\url{http://project-iasis.eu/}}, BigMedilytics - lung cancer pilot\footnote{\url{https://www.bigmedilytics.eu/}}, and P4-LUCAT\footnote{\url{https://p4-lucat.eu/}}. As in Knowlege4COVID-19, the generated KGs include fine-grained representations of publications, and biomedical entities (e.g., drugs, side effects, targets, and interactions); the mapping rules that defined these core concepts have been reused with minor changes. This opens the spectrum of possibilities of reusability and adoption, and puts in perspective the relevance of DEs where KG creation is defined declaratively through mapping rules. 
\section{Conclusions and Future Work}
\label{sec:conclusion}
This paper addresses the problem of providing an integrated view of heterogeneous sources of COVID-19 data. Following the reference architecture of networks of knowledge-driven data ecosystem (by Geisler and Vidal et al.~\cite{abs-2105-09312}), we presented the Knowledge4COVID-19 framework as a data ecosystem (DE) where mappings among data sources and a unified schema are described in terms of RML.
The Knowledge4COVID-19 DE uses the SDM-RDFizer to execute the RML mappings and create the Knowledge4COVID-19 KG. Tasks of Natural Language Processing enable recognizing relevant entities and predicates in the text describing drug-drug interactions and side effects. Additionally, a deductive system and KG predictive models allow the discovery and prediction of patterns to explain the impact of drug-drug interactions on treatment effectiveness and toxicity.
As a result, the Knowledge4COVID-19 KG comprises factual statements about drugs, adverse events, and drug-drug interactions harvested from COVID-19 data sources and scientific publications. 

A repertoire of Web APIs over the Knowledge4COVID-19 KG is made available. They enable exploring entities through their connections and discovering associations to enhance understanding of a SARS-CoV-2 infection and its progression. Thus, Knowledge4COVID-19 broadens the portfolio of semantic web technologies and provides the basis for developing interpretable analytical methods. In the future, we plan to connect the Knowledge4COVID-19 KG to other KGs that maintain COVID-19 related data. Additionally, we would like to extend the KG clinical data about COVID-19 patients and empower Knowledge4COVID-19 DE with the capacity of detecting patterns that can explain the correlation between survival, drug interactions, and adverse events.

\section{Acknowledgements}
\label{sec:acknowledgement}

\noindent This work has been supported by the EU H2020 RIA funded projects iASiS with grant agreement No. 727658, P4-LUCAT with GA No. 53000015, BigMedilytics with GA No. 780495, CLARIFY with GA No. 875160, and Federal Ministry for Economic Affairs and Energy of Germany in the project CoyPu (No 01MK21007[A-L]). Maria-Esther Vidal is also partially supported by Leibniz Association in the program "Leibniz Best Minds: Programme for Women Professors", project TrustKG-Transforming Data in Trustable Insights with grant P99/2020. Knowledge4COVID-19 was originally a project of the Pan-European hackathon \#EUvsVirus in April 2020.

%\section*{References}

\bibliographystyle{unsrtnat}
\bibliography{mybibfile}

\appendix

\section{KG Exploration API queries}
All the following queries are available on our GitHub repository.\footnote{\url{https://github.com/SDM-TIB/Knowledge4COVID-19/blob/main/Exploration-API/SPARQL/README.md}}
\subsection{Publications related to drugs}
\label{appendix:publications}
    \begin{lstlisting}[style=sparql,basicstyle=\footnotesize]
PREFIX k4covid: <http://research.tib.eu/covid-19/vocab/>
PREFIX k4covide: <http://research.tib.eu/covid-19/entity/>
SELECT DISTINCT ?pub ?year ?journal ?title ?url ?drug ?drugLabel where {
        ?drug a k4covid:Drug.
        ?drug k4covid:hasCUIAnnotation ?drugCUI.
        Filter(?drugCUI in (k4covide:C0031623,
        k4covide:C0751995,
        k4covide:C0030106))
        ?drugCUI k4covid:annLabel ?drugLabel.
        ?ann a k4covid:ConceptAnnotation.
        ?ann k4covid:hasSemanticAnnotation ?semAnn.
        ?semAnn k4covid:hasCUIAnnotation ?drugCUI.
        ?ann k4covid:annotates ?pub.
        ?pub <http://purl.org/dc/terms/title> ?title.
        ?pub k4covid:year ?year.
        ?pub k4covid:journal ?journal.
        ?pub k4covid:externalLink ?url.
    }
    \end{lstlisting}
    
\subsection{Drug-Drug Interactions (DDI)}
\subsubsection{Get Interactions of a Drug}
\label{appendix:interactions_drug}
    \begin{lstlisting}[style=sparql,basicstyle=\footnotesize]
PREFIX k4covid: <http://research.tib.eu/covid-19/vocab/>
PREFIX k4covide: <http://research.tib.eu/covid-19/entity/>
SELECT DISTINCT ?effectorDrugLabel ?affectdDrugLabel ?effect AS ?effectLabel  ?impactLabel   WHERE {  
    ?interaction k4covid:precipitantDrug ?effectorDrugCUI.
    ?interaction k4covid:objectDrug ?affectdDrugCUI.
    ?effectorDrugCUI k4covid:annLabel ?effectorDrugLabel.
    ?affectdDrugCUI k4covid:annLabel ?affectdDrugLabel.
    ?interaction k4covid:effect ?effectCUI.
    ?effectCUI k4covid:annLabel  ?effect.
    ?interaction k4covid:impact ?impactLabel.                                   
FILTER(?affectdDrugCUI in (k4covide:C0000970))}
    \end{lstlisting}

\subsubsection{Get all the interaction among the provided Drugs}
\label{appendix:interactions_drugs}
    \begin{lstlisting}[style=sparql,basicstyle=\footnotesize]
PREFIX k4covid: <http://research.tib.eu/covid-19/vocab/>
PREFIX k4covide: <http://research.tib.eu/covid-19/entity/>
SELECT * {
{SELECT DISTINCT ?effectorDrugLabel ?affectdDrugLabel ?effect AS ?effectLabel ?impactLabel  WHERE {
    ?interaction k4covid:precipitantDrug k4covide:C0000970.
    ?interaction k4covid:objectDrug k4covide:C0028978.
    k4covide:C0000970 k4covid:annLabel ?effectorDrugLabel.
    k4covide:C0028978 k4covid:annLabel ?affectdDrugLabel.                 
    ?interaction k4covid:effect ?effectCUI.
    ?effectCUI k4covid:annLabel ?effect.
    ?interaction k4covid:impact ?impactLabel.     
}} UNION 
{SELECT DISTINCT ?effectorDrugLabel ?affectdDrugLabel ?effect AS ?effectLabel ?impactLabel  WHERE {  
    ?interaction k4covid:precipitantDrug k4covide:C0028978.
    ?interaction k4covid:objectDrug k4covide:C0000970.
    k4covide:C0028978 k4covid:annLabel ?effectorDrugLabel.
    k4covide:C0000970 k4covid:annLabel ?affectdDrugLabel.                                        
    ?interaction k4covid:effect ?effectCUI.
    ?effectCUI k4covid:annLabel ?effect.
    ?interaction k4covid:impact ?impactLabel.
}}}                                     

    \end{lstlisting}

\subsection{Predicted Drug-Drug Interactions}
\subsubsection{Get the predicted interactions of a Drug}
\label{appendix:predicted_interactions_drug}
    \begin{lstlisting}[style=sparql,basicstyle=\footnotesize]
PREFIX k4covid: <http://research.tib.eu/covid-19/vocab/>
PREFIX k4covide: <http://research.tib.eu/covid-19/entity/>
SELECT DISTINCT ?effectorDrugLabel ?affectdDrugLabel ?confidence ?provenance WHERE {  
    ?interaction a k4covid:DrugDrugPrediction.
    ?interaction k4covid:hasInteractingDrug ?effectorDrug.
    ?interaction k4covid:hasInteractingDrug ?affectedDrug.
    FILTER (?effectorDrug != ?affectedDrug)
    ?affectedDrug k4covid:hasCUIAnnotation ?affectdDrugCUI.
    ?effectorDrug k4covid:hasCUIAnnotation ?effectorDrugCUI.
    ?effectorDrugCUI k4covid:annLabel ?effectorDrugLabel.
    ?affectdDrugCUI k4covid:annLabel ?affectdDrugLabel.
    ?interaction k4covid:confidence ?confidence.
    ?interaction k4covid:predictionMethod ?provenance.                                
FILTER(?affectdDrugCUI in (k4covide:C0000970))}
    \end{lstlisting}

\subsubsection{Get all the interaction among the provided Drugs}
\label{appendix:predicted_interactions_drugs}
    \begin{lstlisting}[style=sparql,basicstyle=\footnotesize]
PREFIX k4covid: <http://research.tib.eu/covid-19/vocab/>
PREFIX k4covide: <http://research.tib.eu/covid-19/entity/>
SELECT * {
{SELECT DISTINCT ?effectorDrugLabel ?affectdDrugLabel ?confidence ?provenance  WHERE {
    ?interaction k4covid:hasInteractingDrug ?effectorDrug.
    ?interaction k4covid:hasInteractingDrug ?affectedDrug.
    FILTER (?effectorDrug != ?affectedDrug)
    ?effectorDrug k4covid:hasCUIAnnotation k4covide:C0995188.
    ?affectedDrug k4covid:hasCUIAnnotation k4covide:C0765273.
    k4covide:C0000970 k4covid:annlabel ?effectorDrugLabel.
    k4covide:C0009214 k4covid:annlabel ?affectdDrugLabel.
    ?interaction k4covid:confidence ?confidence.
    ?interaction k4covid:predictionMethod ?provenance.       
}} UNION 
{SELECT DISTINCT ?effectorDrugLabel ?affectdDrugLabel ?confidence ?provenance  WHERE {  
    ?interaction k4covid:hasInteractingDrug ?effectorDrug.
    ?interaction k4covid:hasInteractingDrug ?affectedDrug.
    FILTER (?effectorDrug != ?affectedDrug)
    ?effectorDrug k4covid:hasCUIAnnotation k4covide:C0765273.
    ?affectedDrug k4covid:hasCUIAnnotation k4covide:C0995188.
    k4covide:C0009214 k4covid:annLabel ?effectorDrugLabel.
    k4covide:C0000970 k4covid:annLabel ?affectdDrugLabel.
    ?interaction k4covid:confidence ?confidence.
    ?interaction k4covid:predictionMethod ?provenance.    
}}}                                     

    \end{lstlisting}

\end{document}